# COMPARISONS FOR BACKWARD STOCHASTIC DIFFERENTIAL EQUATIONS ON MARKOV CHAINS AND RELATED NO-ARBITRAGE CONDITIONS

By Samuel N. Cohen and Robert J. Elliott

*University of Adelaide and University of Adelaide and University of Calgary*

Most previous contributions to BSDEs, and the related theories of nonlinear expectation and dynamic risk measures, have been in the framework of continuous time diffusions or jump diffusions. Using solutions of BSDEs on spaces related to finite state, continuous time Markov chains, we develop a theory of nonlinear expectations in the spirit of [Dynamically consistent nonlinear evaluations and expectations (2005) Shandong Univ.]. We prove basic properties of these expectations and show their applications to dynamic risk measures on such spaces. In particular, we prove comparison theorems for scalar and vector valued solutions to BSDEs, and discuss arbitrage and risk measures in the scalar case.

**1. Introduction.** Most previous contributions to backward stochastic differential equations (BSDEs), and the related theories of nonlinear expectation and dynamic risk measures, have been in the framework of continuous time diffusions or jump diffusions. The setting has usually been scalar. In this paper a situation is considered where randomness is generated by a continuous time, finite state Markov chain. A corresponding discrete time theory is covered in the paper [5]. In this paper both scalar and vector comparison theorems are proved in this framework. Following [15] this gives rise to a theory of nonlinear expectations. The proofs and results are different from those in the diffusion case.

Consider a continuous time, finite state Markov chain $X = \{X_t, t \in [0, T]\}$. Without loss of generality, we identify the states of this process with the standard unit vectors $e_i$ in $\mathbb{R}^N$ where $N$ is the number of states of the chain.

We consider stochastic processes defined on the filtered probability space $(\Omega, \mathcal{F}, \{\mathcal{F}_t\}, \mathbb{P})$ where $\{\mathcal{F}_t\}$ is the completed natural filtration generated by the $\sigma$-fields $\mathcal{F}_t = \sigma(\{X_u, u \leq t\}, F \in \mathcal{F}_T : \mathbb{P}(F) = 0)$, and $\mathcal{F} = \mathcal{F}_T$. Note that, as $X$ is a right-continuous jump process with distinct jumps, this filtration is right continuous. If $A_t$ denotes the rate matrix for $X$ at time $t$, then this chain has the representation

$$(1.1) \qquad X_t = X_0 + \int_{]0,t]} A_u X_{u-} \, du + M_t,$$

where $M_t$ is a martingale (see Appendix B of [9]). As in [4], we know that the predictable quadratic covariation matrix of this martingale, $\langle M, M \rangle_t$, has the representation

$$(1.2) \quad \langle M, M \rangle_t = \int_{]0,t]} (\mathrm{diag}(A_u X_{u-}) - A_u \, \mathrm{diag}(X_{u-}) - \mathrm{diag}(X_{u-}) A_u^*) \, du.$$

In an earlier paper, [4], we considered pairs, $(Y, Z)$, $Z_t \in \mathbb{R}^{K \times N}$, adapted and left continuous, $Y_t \in \mathbb{R}^K$ adapted and cádlág which are solutions to equations of the form

$$(1.3) \qquad Y_t - \int_{]t,T]} F(\omega, u, Y_{u-}, Z_u) \, du + \int_{]t,T]} Z_u \, dM_u = Q.$$

Here $Q$ is an $\mathcal{F}_T$ measurable, $\mathbb{R}^K$ valued, $\mathbb{P}$-square integrable random variable, and $F$ is a map $\Omega \times [0,T] \times \mathbb{R}^K \times \mathbb{R}^{K \times N} \to \mathbb{R}^K$ which is progressively measurable.

Let $\psi_t$ be the nonnegative definite matrix,

$$(1.4) \qquad \psi_t := \mathrm{diag}(A_t X_{t-}) - A_t \, \mathrm{diag}(X_{t-}) - \mathrm{diag}(X_{t-}) A_t^*$$

and

$$(1.5) \qquad \|Z\|_{X_{t-}}^2 := \mathrm{Tr}(Z \psi_t Z^*).$$

Then $\| \cdot \|_{X_{t-}}$ defines a (stochastic) seminorm, with the property that

$$(1.6) \qquad \mathrm{Tr}(Z_t \, d\langle M, M \rangle_t Z_t^*) = \|Z_t\|_{X_{t-}}^2 \, dt.$$

We have shown the following result in [4].

THEOREM 1.1. *For $Q \in L^2(\mathcal{F}_T)$, if $F$ is Lipschitz continuous, in the sense that there exists $c \geq 0$ such that, for any $Z^1, Z^2, Y^1, Y^2$ square integrable and of appropriate dimension,*

$$\|F(t, Y_{u-}^1, Z_t^1) - F(u, Y_{t-}^2, Z_t^2)\| \leq c(\|Z_t^1 - Z_t^2\|_{X_{t-}} + \|Y_{t-}^1 - Y_{t-}^2\|)$$

$dt \times \mathbb{P}$-a.s., then there exists a solution $(Y, Z)$ to (1.3), such that

$$E\left[\int_{]0,T]} \|Y_t\|^2 \, du\right] < +\infty,$$

$$E\left[\int_{]0,T]} \|Z_t\|^2_{X_{t-}} \, du\right] < +\infty,$$

and this solution is the unique such solution, up to indistinguishability for $Y$ and equality $d\langle M, M\rangle_t \times \mathbb{P}$-a.s. for $Z$.

In this paper, we shall develop properties and applications of these solutions.

Note that, where appropriate, we shall denote by $\leq$, $\geq$, etc. an inequality holding simultaneously on all components of a vector quantity.

**2. Extension to stopping times.** We first extend the results of [4] to include the case where $T$, the terminal time of our BSDE, is an essentially bounded stopping time. By this we mean that $T$ is a stopping time and there exists a real value $T^{\max}$ such that $\mathbb{P}(T > T^{\max}) = 0$.

Assume that $F(\omega, t, Y_{t-}, Z_t)$ is well defined and Lipschitz continuous $dt \times \mathbb{P}$-a.s. on $[0, T^{\max}]$. Then from Theorem 1.1, there is a unique solution of the BSDE,

$$(2.1) \quad Y_t - \int_{]t,T^{\max}]} I_{u \leq T} F(\omega, u, Y_{u-}, Z_u) \, du + \int_{]t,T^{\max}]} Z_u \, dM_u = Q,$$

for any $Q \in L^2(\mathcal{F}_{T^{\max}})$; this of course includes all $Q \in L^2(\mathcal{F}_T)$, as $\mathbb{P}(T \leq T^{\max}) = 1$. For $\mathcal{F}_T$ measurable $Q$, we evaluate (2.1) at $T$, and take an $\mathcal{F}_T$ conditional expectation. This gives

$$Y_T + E\left[\int_{]T,T^{\max}]} Z_u \, dM_u \Big| \mathcal{F}_T\right] = Q$$

$\mathbb{P}$-a.s., and as $M$ is a martingale we see $Y_T = Q$ $\mathbb{P}$-a.s. It follows from this that $\int_{]T,T^{\max}]} Z_u \, dM_u = 0$ $\mathbb{P}$-a.s., hence $Z_t = 0$, $d\langle M, M\rangle_t \times \mathbb{P}$ almost surely, for all $T < t \leq T^{\max}$. Consequently, we can rewrite our BSDE as

$$(2.2) \quad Y_t - \int_{]t \wedge T,T]} F(\omega, u, Y_{u-}, Z_u) \, du + \int_{]t \wedge T,T]} Z_u \, dM_u = Q,$$

which is a useful generalization of (1.3).

We shall restrict ourselves to deterministic $T$ in the following; however, it is clear that, given appropriate modifications, the results stated could easily be modified to remain valid when $T$ is an essentially bounded stopping time.

**3. Basic theorems.** Before developing specific applications of these processes, we establish the following results. The methods used are based on those in [7], where similar results are proven for BSDEs on spaces related to Brownian motions.

We shall henceforth assume that $F$ is $\mathbb{P}$-a.s. left continuous in $t$, is Lipschitz continuous as in Theorem 1.1 and satisfies

$$E\left[\int_{]0,T]} \|F(\omega,t,Y_{t-},Z_t)\|^2 \, dt\right] < +\infty$$

for all $Y$, $Z$ bounded as in Theorem 1.1. Such a driver will be called *standard*. If also $Q \in L^2(\mathcal{F}_T)$, then the pair $(F,Q)$ will be called standard.

We shall assume that the rate matrix $A$ of our chain is left continuous, and there is an $0 < \varepsilon_r < 1$ such that it satisfies

$$(3.1) \qquad e_i^* A_t e_j \in [\varepsilon_r, 1/\varepsilon_r] \cup \{0\}$$

$dt$-a.s., for all $i$ and $j$, $i \neq j$. This assumption is trivially satisfied if the chain $X$ is time-homogenous, and essentially states that we shall not consider chains with positive transition rates unbounded or arbitrarily close to zero.

3.1. *Various lemmas.* Throughout this section, **1** will denote a column vector of appropriate dimension with all components equal to one.

We first note that, from the basic properties of stochastic integrals (see, e.g., [14], page 28), the following isometry holds.

LEMMA 3.1. *For any predictable (matrix) process $Z$ (of appropriate dimension) any $s < t$,*

$$E\left[\left\|\int_{]s,t]} Z_u \, dM_u\right\|^2\right] = E\left[\int_{]s,t]} \|Z_u\|_{X_{t-}}^2 \, du\right].$$

PROOF. Let $\langle \cdot, \cdot \rangle$ denote the predictable quadratic covariation. Expanding the norms in terms of traces and products, we wish to show

$$(3.2) \qquad \begin{aligned} E\,\mathrm{Tr}&\left[\left(\int_{]s,t]} Z_u \, dM_u\right)\left(\int_{]s,t]} Z_u \, dM_u\right)^*\right] \\ &= E\left[\int_{]s,t]} \mathrm{Tr}[Z_u \, d\langle M, M\rangle_u (Z_u)^*]\right]. \end{aligned}$$

If $\int_{]s,t]} Z_u \, dM_u$ is square integrable, then

$$\left(\int_{]s,t]} Z_u \, dM_u\right)\left(\int_{]s,t]} Z_u \, dM_u\right)^* - \left\langle \int_{]s,\cdot]} Z_u \, dM_u, \int_{]s,\cdot]} Z_u \, dM_u\right\rangle_t$$

is a uniformly integrable martingale (see [14], page 38), and

$$\left\langle \int_{]s,\cdot]} Z_u \, dM_u, \int_{]s,\cdot]} Z_u \, dM_u \right\rangle_t = \int_{]s,t]} Z_u \, d\langle M, M \rangle_u (Z_u)^*$$

(by [14], page 48, Theorem 4.4). Combining these equations and taking a trace and an expectation gives the result in this case.

If $\int_{]s,t]} Z_u \, dM_u$ is not square integrable, then, as both sides of (3.2) are nonnegative, they both equal $+\infty$. The result follows. $\square$

LEMMA 3.2. *For* $(F, Q)$ *standard, if* $Y$ *is the solution to* (1.3) *given by Theorem 1.1, then* $Y$ *satisfies*

$$\sup_{t \in [0,T]} E[\|Y_t\|^2] < +\infty.$$

PROOF. We have

$$\sup_{t \in [0,T]} E[\|Y_t\|^2] = \sup_{t \in [0,T]} E\left[ \left\| Q + \int_{]t,T]} F(\omega, u, Y_{u-}, Z_u) \, du - \int_{]t,T]} Z_u \, dM_u \right\|^2 \right]$$

$$\leq 2E[\|Q\|^2] + 4E\left[ \int_{]0,T]} \|F(\omega, u, Y_{u-}, Z_u)\|^2 \, du \right]$$

$$\quad + 4 \sup_{t \in [0,T]} E\left[ \left\| \int_{]t,T]} Z_u \, dM_u \right\|^2 \right]$$

$$\leq 2E[\|Q\|^2] + 4E\left[ \int_{]0,T]} \|F(\omega, u, Y_{u-}, Z_u)\|^2 \, du \right]$$

$$\quad + 4E\left[ \int_{]0,T]} \|Z_u\|_{X_{u-}}^2 \, du \right].$$

As $Q \in L^2(\mathcal{F}_T)$, $F$ is standard and $E[\int_{]0,T]} \|Z_u\|_{X_{u-}}^2 \, du]$ is finite by the conditions of Theorem 1.1; this gives the desired inequality. $\square$

LEMMA 3.3. *For all* $t$, *the matrix* $\psi_t$ *is symmetric and bounded, and has all row and column sums equal to zero. The matrix given by its Moore–Penrose inverse,* $\psi_t^+$, *is also bounded and has all row and column sums equal to zero.*

PROOF. As $A_t$ is a rate matrix, $A_t^* \mathbf{1} = 0$. Hence, from (1.4),

$$\psi_t \mathbf{1} = \operatorname{diag}(A_t X_{t-}) \mathbf{1} - A_t \operatorname{diag}(X_{t-}) \mathbf{1} - \operatorname{diag}(X_{t-}) A_t^* \mathbf{1}$$

$$= A_t X_{t-} - A_t X_{t-} - 0$$

$$= 0 \in \mathbb{R}^N,$$

that is, the rows of $\psi_t$ all sum to zero, and so $\psi_t$ is singular.

The symmetry of $\psi_t$ is trivial, which implies the columns of $\psi_t$ also all sum to zero. For clarity, an example of $\psi_t$ is presented for $N = 4$ in the Appendix.

For any matrix with real components, the Moore–Penrose inverse exists. We first note that, if and only if $e_j^* A_t X_{t-} = 0$, the $j$th row and column of $\psi_t$ will be zero. Also, if we exclude the row and column indicated by $X_{t-}$, $\psi_t$ is a diagonal matrix. Therefore, if $\psi_t$ has $m$ rows and columns of zeros, the rank of $\psi_t$ will be $N - 1 - m$ (unless $m = N$, in which case it will be zero). Let $\tilde{\psi}_t$ denote $\psi_t$ excluding the row and column corresponding to $X_{t-}$ and any rows and columns of zeros. Using the results of [6], Theorem 6.1, we can express $\psi_t^+$ as the inverse of $\tilde{\psi}_t$, pre- and post-multiplied by bounded matrices with row and column sums of zero, as appropriate. As $\tilde{\psi}_t$ is a diagonal matrix, its inverse is bounded by (3.1). Hence $\psi_t^+$ is bounded, and has all row and column sums equal to zero. □

LEMMA 3.4. *At each time $t$, for all $j$ such that $e_j^* A_t X_{t-} \neq 0$,*

$$\psi_t^+ \psi_t (e_j - X_{t-}) = \psi_t \psi_t^+ (e_j - X_{t-}) = e_j - X_{t-}.$$

*Furthermore, the vectors $(e_j - X_{t-})$ form a basis for the range of the projections, $\psi_t^+ \psi_t$ and $\psi_t \psi_t^+$.*

PROOF.  Again we use the results of [6], in particular Theorems 6.1, 7.1 and 7.2. By Cohen, Elliott and Pearce [6], Theorem 6.1, we know that $\psi_t^+$ is a true inverse for $\psi_t$, and $(\psi_t^+)^+$ is a true inverse for $\psi_t^+$, within the class of matrices with both row and column sums zero and with rows and columns of zeros in the same rows and columns as $\psi_t$. Therefore $\psi_t = (\psi_t^+)^+$. The desired equation is then simply the result of [6], Theorem 7.1. The remaining statement is then simply an application of [6], Theorem 7.2. □

LEMMA 3.5. *For any standard driver $F$, for any $Y$ and $Z$, up to indistinguishability,*

$$F(\omega, t, Y_{t-}, Z_t) = F(\omega, t, Y_{t-}, Z_t \psi_t \psi_t^+)$$

*and*

$$\int_{]0,t]} Z_u \, dM_u = \int_{]0,t]} Z_u \psi_u \psi_u^+ \, dM_u.$$

*Therefore, it is no loss of generality to assume that $Z = Z \psi \psi^+$.*

PROOF. Recall that, from the definition of the Moore–Penrose inverse, $\psi_t \psi_t^+ \psi_t = \psi_t$. Hence, for any $t$, $\mathbb{P}$-a.s.,

$$\int_{]0,t]} \|Z_u - Z_u \psi_u \psi_u^+\|_{X_{t-}}^2 \, du$$

$$= \int_{]0,t]} \operatorname{Tr}([Z_u - Z_u \psi_u \psi_u^+] \, d\langle M, M \rangle_u [Z_u - Z_u \psi_u \psi_u^+]^*)$$

$$= \int_{]0,t]} \operatorname{Tr}([Z_u - Z_u \psi_u \psi_u^+] \psi_t [Z_u - Z_u \psi_u \psi_u^+]^*) \, dt$$

$$= 0.$$

By the Lipschitz continuity of $F$ this implies

$$E\left[\int_{]0,t]} \|F(\omega, u, Y_{u-}, Z_u) - F(\omega, u, Y_{u-}, Z_u \psi_u \psi_u^+)\|^2 \, du\right]$$

$$\leq c^2 E\left[\int_{]0,t]} \|Z_u - Z_u \psi_u \psi_u^+\|_{X_{u-}}^2 \, du\right]$$

$$= 0.$$

As $F$ is $\mathbb{P}$-a.s. left continuous in $t$, and $Y_{t-}$ and $Z$ are left continuous processes, we use the result of [8], Lemma 2.21, to show that

$$F(\omega, t, Y_{t-}, Z_t) = F(\omega, t, Y_{t-}, Z_t \psi_t \psi_t^+)$$

up to indistinguishability.

By Lemma 3.1,

$$E\left[\left\|\int_{]0,t]} (Z_u - Z_u \psi_u \psi_u^+) \, dM_u\right\|^2\right] = E\left[\int_{]0,t]} \|Z_u - Z_u \psi_u \psi_u^+\|_{X_{t-}}^2 \, du\right] = 0$$

and so for all $t$, $\mathbb{P}$-a.s.,

$$\int_{]0,t]} (Z_u - Z_u \psi_u \psi_u^+) \, dM_u = 0.$$

As this process is a stochastic integral, it is càdlàg (see [14], Theorem 4.31), and therefore we can again use [8], Lemma 2.21, to show the stochastic integrals are equal up to indistinguishability. □

LEMMA 3.6. $\|Z_t\|_{X_{t-}} = 0$ only if $Z_t \psi_t \psi_t^+ = 0$. Hence without loss of generality, $\|Z_t\|_{X_{t-}} = 0$ if and only if $Z_t = 0$.

PROOF. We know that

$$0 = \|Z_t\|_{X_{t-}} = \operatorname{Tr}(Z_t \psi_t Z_t^*).$$

As $\psi_t$ is nonnegative definite, there exists a matrix $P$ such that $\psi_t = PP^*$, the Cholesky decomposition of $\psi_t$. Hence

$$0 = \text{Tr}(Z_t \psi_t Z_t^*) = \text{Tr}(Z_t P (Z_t P)^*),$$

which is true if and only if $Z_t P = 0$. Hence, without loss of generality,

$$Z_t = Z_t \psi_t \psi_t^+ = Z_t P P^* \psi_t^+ = 0. \qquad \square$$

LEMMA 3.7.   *For all $t \in [0, T]$,*

$$(3.3) \qquad \psi_t X_{t-} = -A_t X_{t-} = -\sum_j (e_j^* A_t X_{t-})(e_j - X_{t-}).$$

PROOF.   From (1.4),

$$\psi_t X_{t-} = \text{diag}(A_t X_{t-}) X_{t-} - A_t \, \text{diag}(X_{t-}) X_{t-} - \text{diag}(X_{t-}) A_t^* X_{t-}.$$

As $X_{t-}$ is a standard basis vector in $\mathbb{R}^N$, we have

$$\psi_t X_{t-} = \text{diag}(A_t X_{t-}) X_{t-} - A_t X_{t-} - (\text{diag}(X_{t-}) A_t^*) X_{t-}.$$

Further,

$$\text{diag}(A_t X_{t-}) X_{t-} = (X_{t-}^* A_t X_{t-}) X_{t-} = (\text{diag}(X_{t-}) A_t^*) X_{t-},$$

therefore

$$\psi_t X_{t-} = -A_t X_{t-},$$

establishing the first equality.

We now note that, as $\mathbf{1}^* A_t = 0$, we have

$$X_{t-}^* A_t X_{t-} = -\sum_{j \,:\, X_{t-} \neq e_j} (e_j^* A_t X_{t-})$$

and hence

$$A_t X_{t-} = \sum_{j \,:\, X_{t-} \neq e_j} (e_j^* A_t X_{t-}) e_j - \left[ \sum_{j \,:\, X_{t-} \neq e_j} (e_j^* A_t X_{t-}) \right] X_{t-}$$

$$= \sum_j (e_j^* A_t X_{t-})(e_j - X_{t-}). \qquad \square$$

LEMMA 3.8.   *For processes $Z$ solving (1.3), without loss of generality, the $\| \cdot \|_{X_{t-}}$ norm has two equivalent forms:*

$$\|Z_t\|_{X_{t-}}^2 = \text{Tr}(Z_t \psi_t Z_t^*)$$

$$= \sum_{i,j} (e_j^* A_t X_{t-}) [e_i^* Z_t (e_j - X_{t-})]^2.$$

Proof. Consider the $\|\cdot\|_{X_{t-}}$ norm of $Z_t$. The trace can be calculated as $\mathrm{Tr}(Z_t \psi_t Z_t^*) = \sum_i e_i^* Z_t \psi_t Z_t^* e_i$, and therefore it is clear that we can consider each row $e_t^* Z$ separately from the others. Because of this, we need only to establish the result where $Z$ is a single *row* vector.

For any row vector of the form $v = \sum_{e_j \neq X_{t-}} c_j e_j^*$, it is clear that

$$\mathrm{diag}(X_{t-})v^* = \mathbf{0}$$

and, therefore,

$$
\begin{aligned}
(3.4) \quad \|v\|_{X_{t-}} &= v\psi_t v^* \\
&= v(\mathrm{diag}(A_t X_{t-}) - A_t \mathrm{diag}(X_{t-}) - \mathrm{diag}(X_{t-})A_t^*)v^* \\
&= \left(\sum_{e_j \neq X_{t-}} c_j e_j^*\right)(\mathrm{diag}(A_t X_{t-}))\left(\sum_{e_j \neq X_{t-}} c_j e_j\right) \\
&= \sum_{e_j \neq X_{t-}} (e_j^* A_t X_{t-})c_j^2.
\end{aligned}
$$

We shall use this "linearity" to establish the result.

By Lemma 3.5, without loss of generality, we can write $Z_t = Z_t \psi_t \psi_t^+$. We can therefore write $Z_t$ as a linear combination $Z_t = \sum_j c_j(e_j - X_{t-})^*$, by Lemmas 3.3 and 3.4. For simplicity, we define $c_j = 0$ if $e_j = X_{t-}$.

We define a vector of the form considered above,

$$
\begin{aligned}
(3.5) \quad \tilde{Z}_t &= \sum_{e_j \neq X_{t-}} \left(c_j + \sum_k c_k\right) e_j^* \\
&= \sum_j c_j e_j^* + \left(\sum_k c_k\right)\mathbf{1}^* - \left(\sum_k c_k\right)X_{t-}^*
\end{aligned}
$$

and as $\psi_t$ is a matrix with column sums of zero, this and Lemma 3.4 imply

$$\tilde{Z}_t \psi_t \psi_t^+ = \sum_j c_j e_j^* - \left(\sum_j c_j\right)X_{t-}^* = Z_t.$$

By Lemma 3.4, we then know

$$Z_t(e_j - X_{t-}) = \tilde{Z}_t \psi_t \psi_t^+(e_j - X_{t-}) = \tilde{Z}_t(e_j - X_{t-})$$

and it follows from (3.5) that

$$(3.6) \quad c_j + \sum_k c_k = \tilde{Z}_t(e_j - X_{t-}) = Z_t(e_j - X_{t-})$$

for all $e_j \neq X_{t-}$.

As $Z = \tilde{Z}\psi_t\psi_t^+$ and $\psi_t\psi_t^+\psi_t = \psi_t$, it is clear from (1.5) that

$$
\begin{aligned}
\|Z\|_{X_{t-}}^2 &= \|\tilde{Z}\|_{X_{t-}}^2 \\
&= \sum_{e_j \neq X_{t-}} (e_j^* A_t X_{t-})\Big[c_j + \sum_k c_k\Big]^2 \\
&= \sum_j (e_j^* A_t X_{t-})[Z(e_j - X_{t-})]^2.
\end{aligned}
$$

$\square$

LEMMA 3.9.  *Without loss of generality, the row sums of $Z_t$ and $Z_t\psi_t$ are all zero for all $t$.*

PROOF.   The row sums of $Z$ can be written $Z\mathbf{1}$. As $\psi$ and $\psi^+$ are matrices with all row and column sums equal to zero, $\psi\mathbf{1} = \psi^+\mathbf{1} = 0$ which directly shows, using Lemma 3.5, $Z\mathbf{1} = Z\psi\psi^+\mathbf{1} = 0$ and $Z\psi\mathbf{1} = 0$.   $\square$

LEMMA 3.10.   *Consider a process $Z \in \mathbb{R}^{1 \times N}$ solving a standard, scalar, $(K = 1)$ BSDE of the form of (1.3). Suppose that for a given $t$, $Z_t$ is such that $\|Z_t\|_{X_{t-}} \neq 0$, and, for some $\varepsilon$,*

$$
0 < \varepsilon < \varepsilon_r^{3/2} N^{-3/2}
$$

*with $\varepsilon_r$ as in (3.1), we know*

$$
-\varepsilon\|Z_t\|_{X_{t-}} \leq (e_j^* A_t X_{t-}) Z_t(e_j - X_{t-})
$$

*for all $j$. Then*

$$
Z_t\psi_t X_{t-} \leq -\varepsilon\|Z_t\|_{X_{t-}}.
$$

PROOF.   We know from Lemma 3.8 that

$$
\begin{aligned}
\|Z_t\|_{X_{t-}}^2 &= \mathrm{Tr}(Z_t\psi_t Z_t^*) \\
&= \sum_j (e_j^* A_t X_{t-})[Z_t(e_j - X_{t-})]^2.
\end{aligned}
$$

By (3.1), for all $e_j \neq X_{t-}$, $e_j^* A_t X_{t-} \in [\varepsilon_r, 1/\varepsilon_r] \cup \{0\}$, for some $\varepsilon_r > 0$, and so

$$
\begin{aligned}
(3.7) \qquad \|Z_t\|_{X_{t-}}^2 &\leq \sum_{\{j\,:\,e_j^* A_t X_{t-} > 0\}} \varepsilon_r^{-1}[Z_t(e_j - X_{t-})]^2 \\
&\leq N\varepsilon_r^{-1} \max_{\{j\,:\,e_j^* A_t X_{t-} > 0\}} [Z_t(e_j - X_{t-})]^2.
\end{aligned}
$$

Multiplying both sides by $N^{-1}\varepsilon_r$, this implies there must be an $e_j$ with $e_j^* A_t X_{t-} > 0$ such that

$$\varepsilon_r N^{-1}\|Z_t\|_{X_{t-}}^2 \leq [Z_t(e_j - X_{t-})]^2.$$

As $e_j^* A_t X_{t-} \geq \varepsilon_r$, we have

$$\varepsilon_r^{3/2} N^{-1/2}\|Z_t\|_{X_{t-}} \leq (e_j^* A_t X_{t-})Z_t(e_j - X_{t-}).$$

For all $e_j$ such that $(e_j^* A_t X_{t-})Z_t(e_j - X_{t-}) < 0$, we have, by assumption,

$$-\varepsilon\|Z_t\|_{X_t} \leq (e_j^* A_t X_{t-})Z_t(e_j - X_{t-}) < 0.$$

By Lemma 3.7, we deduce

$$\begin{aligned}
Z_t\psi_t X_{t-} &= -Z_t A_t X_{t-}\\
&= -\sum_j (e_j^* A_t X_{t-})Z_t(e_j - X_{t-})\\
&\leq -\varepsilon_r^{3/2} N^{-1/2}\|Z_t\|_{X_{t-}}^2 + (N-1)\varepsilon\|Z_t\|_{X_{t-}}\\
&\leq -\varepsilon\|Z_t\|_{X_{t-}},
\end{aligned}$$

where the final inequality is from the assumption that $\varepsilon < \varepsilon_r^{3/2} N^{-3/2}$. □

### 3.2. *Linear BSDEs.*

THEOREM 3.11 (Linear BSDEs). *Let $(\alpha, \beta, \gamma)$ be a $du \times \mathbb{P}$-a.s. bounded $(\mathbb{R}^{K \times N}, \mathbb{R}^{K \times K}, \mathbb{R}^K)$ valued predictable process, $\phi$ a predictable $\mathbb{R}^K$ valued process with $E[\int_{[0,T]} \|\phi_t\|^2\, dt] < +\infty$, $T$ a deterministic terminal time and $Q$ a square-integrable $\mathcal{F}_T$ measurable $\mathbb{R}^K$ valued random variable. Then the linear BSDE given by*

$$\tag{3.8} Y_t - \int_{]t,T]} [\phi_u + \beta_u Y_{u-} + \alpha_u Z_u^* \gamma_u]\, du + \int_{]t,T]} Z_u\, dM_u = Q$$

*has a unique square integrable solution $(Y, Z)$ (up to appropriate sets of measure zero). Furthermore, if*

$$\tag{3.9} I + \alpha_s \psi_s^+(e_j - X_{s-})\gamma_s^*$$

*is invertible for all $j$ such that $e_j^* A_s X_{s-} > 0$, except possibly on some evanescent set, then $Y$ is given by the explicit formula*

$$\tag{3.10} Y_t = E\left[\Gamma_t^T Q + \int_{]t,T]} \Gamma_t^u \phi_u\, du \,\Big|\, \mathcal{F}_t\right]$$

*up to indistinguishability. Here $\Gamma_t^s$ is the adjoint process defined for $t \leq s \leq T$ by the forward linear SDE,*

$$\tag{3.11} \Gamma_t^s = I + \int_{]t,s]} \Gamma_t^{u-}[\beta_u\, du + \alpha_u \psi_u^+\, dM_u \gamma_u^*].$$

Before proving this, we establish the following results:

LEMMA 3.12.  *The adjoint process defined by (3.11) will satisfy, for $t \leq r \leq s$,*

$$\Gamma_t^r \Gamma_r^s = \Gamma_t^s.$$

PROOF.  Assume without loss of generality $t < s < T$. Write

$$dV_u = \beta_u \, du + \alpha_u \psi_u^+ \, dM_u \gamma_u^*.$$

Then $\Gamma$ is defined by the forward SDE,

$$\Gamma_t^s = I + \int_{]t,s]} \Gamma_t^{u-} \, dV_u.$$

If $H$ is $\mathcal{F}_t$ measurable in $\mathbb{R}^{N \times N}$, then

$$L = H\Gamma$$

is the unique solution to

$$L_t^s = H + \int_{]t,s]} L_t^{u-} \, dV_u.$$

Suppose $t \leq r$, and define

$$D_t^s = \begin{cases} \Gamma_t^s, & \text{for } t \leq s \leq r, \\ \Gamma_t^r \Gamma_r^s, & \text{for } t \leq r \leq s. \end{cases}$$

Then

$$D_t^s = I + \int_{]t,s]} D_t^{u-} \, dV_u$$

for $t \leq s \leq r$, and

$$D_t^s = \Gamma_t^r + \int_{]r,s]} D_t^{u-} \, dV_u$$

$$= I + \int_{]t,s]} D_t^{u-} \, dV_u$$

for $t \leq r \leq s$. By uniqueness, for $t \leq r \leq s$, this implies $D_t^s = \Gamma_t^s = \Gamma_t^r \Gamma_r^s$, as desired.  □

We now discuss how solutions of first-order matrix differential equations can be expressed as product integrals.

LEMMA 3.13. *Consider a deterministic first-order differential equation of the form*

$$(3.12) \qquad G_s^t = I + \int_{]s,t]} G_s^{u-} H_u \, du.$$

*Here $G$ and $H$ are $K \times K$ matrix-valued functions, $H$ is bounded and Lebesgue integrable and $s < t$. Then (3.12) has an invertible solution, which can be expressed as a product integral, denoted by*

$$G_s^t = \Pi_{]s,t]}(I + H_u \, du).$$

*If $H_t$ has nonnegative entries off the main diagonal for $dt$-almost all $t$, then this solution has all entries nonnegative.*

PROOF. As $H$ is bounded and Lebesgue integrable, we can appeal to the theory of the product integral. Using this,

$$(3.13) \quad G_s^t = \Pi_{]s,t]}(I + H_u \, du) = \lim_{n \to \infty} \prod_{i=0}^{n-1} \left\{ I + \int_{](s_n)_i,(s_n)_{i+1}]} H_u \, du \right\},$$

where, for each $n$, $(s_n)$ is an arbitrary partition of $]s,t]$ into $n$ parts,

$$s = (s_n)_0 < (s_n)_1 < (s_n)_2 < \cdots < (s_n)_n = t,$$

converging, as $n \to \infty$, in the sense that

$$\lim_{n \to \infty} \max_i |(s_n)_i - (s_n)_{i+1}| = 0,$$

$\prod$ indicates products taken sequentially on the right and the limit is taken in the matrix norm $\mathrm{Tr}(G_s^t(G_s^t)^*)$. This is called the product integral of $H$, exists by [13], Theorem 1 and solves the integral equation, (3.12) by [13], Theorem 5.

$G_s^t$ has an inverse, given by

$$(3.14) \qquad (G_s^t)^{-1} = \lim_{n \to \infty} \prod_{i=n-1}^{0} \left\{ I - \int_{](s_n)_i,(s_n)_{i+1}]} H_u \, du \right\}$$

as noted in [12], page 134.

Finally, if $H$ has nonnegative entries off the main diagonal $dt$-a.e., as $H$ is bounded, for sufficiently large $n$, $\int_{](s_n)_i,(s_n)_{i+1}]} H_u \, du$ has all diagonal entries greater than $-1$ for all $i$. Therefore,

$$I + \int_{](s_n)_i,(s_n)_{i+1}]} H_u \, du$$

is nonnegative for all $i$. Consequently, the product in (3.13) must be nonnegative. As the set of matrices with nonnegative components is closed, the limit $\Pi_{]s,t]}(I + H_u \, du)$ will also have all components nonnegative. □

LEMMA 3.14. *Let $t$ be a time at which $X$ jumps, that is, $\Delta X_t \neq 0$. Then*

$$\mathbb{P}(\Delta X_t = e_j - X_{t-} \text{ for some } j \text{ with } e_j^* A_t X_{t-} > 0) = 1.$$

PROOF. Recall that $e_j^* A_u e_i$ is left continuous and lies in $[\varepsilon_r, \varepsilon_r^{-1}] \cup \{0\}$. If for some time $u$ we have $e_j^* A_u e_i = 0$, then there exist $r < u \leq s$ such that $e_j^* A_u e_i = 0$ for all $u \in \, ]r, s]$. As $A$ is left continuous and $X$ is cádlàg, $A$ and $X$ have countably many discontinuities, and therefore, for each $e_i, e_j$, there is a countable set of stopping times $r_n, s_n$ such that $X_{r_n} = e_i$ and

$$e_j^* A_u e_i = 0 \qquad \text{for all } u \in \, ]r_n, s_n].$$

It is then clear that, for a given pair $e_i, e_j$,

$$\{u : e_j^* A_u X_{u-} = 0 \text{ and } X_{u-} = e_i\} = \bigcup_n \, ]r_n, s_n].$$

For a given $n$, let

$$\tau = \min\{t > r_n : \Delta X_t \neq 0\} \wedge s_n.$$

Then

$$E[X_\tau | \mathcal{F}_{r_n}] = X_{r_n} + E\left[ \int_{]r_n, \tau]} A_u X_{u-} \, du \Big| \mathcal{F}_{r_n} \right]$$

and hence

$$E[e_j^* \Delta X_\tau] = E\left[ E\left[ \int_{]r_n, \tau]} e_j^* A_u X_{u-} \, du \Big| \mathcal{F}_{r_n} \right] \right] = 0.$$

It is clear that the left-hand side of this equation is positive if and only if $\mathbb{P}(\Delta X_\tau = e_j - e_i) > 0$. Hence we have shown that

$$\mathbb{P}(\Delta X_t = e_j - e_i \text{ for some } t \in \, ]r_n, s_n]) = 0.$$

Taking a sum over all $i, j$ and the countable index $n$ gives the desired result. □

LEMMA 3.15. *Suppose that for all $s \in \, ]t, T]$,*

$$I + \alpha_s \psi_s^+ (e_j - X_{s-}) \gamma_s^*$$

*is invertible for all $j$ such that $e_j^* A_s X_{s-} > 0$, except possibly on some evanescent set. Then the adjoint process $\Gamma_t^s$ defined by (3.11) is invertible (except possibly on this evanescent set).*

PROOF. By the definition of $M$ in (1.1), we can rewrite (3.11) as

$$
\begin{aligned}
(3.15) \quad \Gamma_t^s = I &+ \int_{]t,s]} \Gamma_t^u [\beta_u - \alpha_u \psi_u^+ A_u X_{u-} \gamma_u^*] \, du \\
&+ \sum_{t < u \le s} \Gamma_t^{(u-)} \alpha_u \psi_u^+ \Delta X_u \gamma_u^*.
\end{aligned}
$$

When $\Delta X_u = 0$, (3.15) is of the form of a classical, deterministic, linear matrix differential equation, with nonsingular solution $\Gamma_{s_0}^s$ given by the analogue of (3.13). Therefore, if two consecutive jump times are $s_0, s_1$, for $s \in [s_0, s_1[$ we have

$$
\Gamma_t^s = \Gamma_t^{s_0} \Pi_{]s_0, s]} \{ I + [\beta_u - \alpha_u \psi_u^+ A_u X_{u-} \gamma_u^*] \, du \} = \Gamma_t^{s_0} \Gamma_{s_0}^s
$$

by Lemma 3.13. If $\Gamma_t^{s_0}$ is invertible, then

$$
(3.16) \quad (\Gamma_t^s)^{-1} = (\Gamma_{s_0}^s)^{-1} (\Gamma_t^{s_0})^{-1}
$$

exists.

At each jump time, (3.15) implies we have

$$
\Delta \Gamma_t^s = \Gamma_t^{(s-)} \alpha_s \psi_s^+ \Delta X_s \gamma_s^*.
$$

Hence

$$
\Gamma_t^s = \Gamma_t^{(s-)} (I + \alpha_s \psi_s^+ \Delta X_s \gamma_s^*).
$$

By Lemma 3.14, at each jump time $s$,

$$
\mathbb{P}(\Delta X_s = e_j - X_{s-} \text{ for some } j \text{ with } e_j^* A_s X_{s-} > 0) = 1.
$$

By assumption, $I + \alpha_s \psi_s^+ (e_j - X_{s-}) \gamma_s^*$ is invertible for all such $j$ (up to evanescence), and

$$
(3.17) \quad (\Gamma_t^s)^{-1} = (I + \alpha_s \psi_s^+ \Delta X_s \gamma_s^*)^{-1} (\Gamma_t^{(s-)})^{-1}.
$$

The process $X$ almost surely has finitely many jumps in $[0, T]$. Through a process of induction using the starting value $\Gamma_t^t = I$ and (3.16) and (3.17), we can conclude $\Gamma_t^s$ is invertible (up to evanescence). $\quad\square$

THEOREM 3.16. *Suppose $\beta_u - \alpha_u \psi_u^+ A_u X_{u-} \gamma_u^*$ has all nonnegative components off the main diagonal $\mathbb{P} \times du$-a.s., for $u \in ]t, T]$, and, except possibly on some evanescent subset of $\Omega \times ]t, T]$,*

$$
I + \alpha_u \psi_u^+ (e_j - X_{u-}) \gamma_u^*
$$

*has nonnegative components for all $j$ such that $e_j^* A_u X_{u-} > 0$. Then the adjoint process $\Gamma_t^s$ has all entries nonnegative for all $s \in ]t, T]$, up to evanescence.*

PROOF.   As above, if two consecutive jump times are $s_0, s_1$, for $s \in [s_0, s_1[$ we have from (3.15),

$$\Gamma_t^s = \Gamma_t^{s_0} \Pi_{]s_0, s]} \{ I + [\beta_u - \alpha_u \psi_u^+ A_u X_{u-} \gamma_u^*] \, du \}.$$

By assumption, $[\beta_u - \alpha_u \psi_u^+ A_u X_{u-} \gamma_u^*] \, du$ has nonnegative components off the main diagonal, and therefore by Lemma 3.13, $\Gamma_{s_0}^s$ has nonnegative components. The product of matrices with nonnegative components has nonnegative components, so if $\Gamma_t^{s_0}$ has nonnegative components, $\Gamma_t^s$ has nonnegative components.

At each jump time, we have from (3.15) that $\Delta \Gamma_t^s = \Gamma_t^{(s-)} \alpha_s \psi_s^+ \Delta X_s \gamma_s^*$. This implies

$$\Gamma_t^s = \Gamma_t^{(s-)} (I + \alpha_s \psi_s^+ \Delta X_s \gamma_s^*)$$

and the term in parentheses has all nonnegative components by our assumption and the argument of Lemma 3.15 regarding the values of $\Delta X_s$.

The process $X$ almost surely has finitely many jumps in $[0, T]$, and therefore through a process of induction using the starting value $\Gamma_t^t = I$, we conclude that $\Gamma_t^s$ has nonnegative entries up to evanescence.   $\square$

COROLLARY 3.17.   *The conditions of Theorem 3.16 are necessary for $\Gamma_t^s$ to have all entries nonnegative for all $s$ and $t$.*

PROOF.   First consider the case when $s$ is not a jump time. Then, as noted by [3], page 176, for small $\delta > 0$,

$$\Gamma_{s-\delta}^s = I + [\beta_{s-\delta} - \alpha_{s-\delta} \psi_{s-\delta}^+ A_{s-\delta} X_{(s-\delta)-} \gamma_u^*] \delta + O(\delta^2).$$

Hence if $[\beta_{s-\delta} - \alpha_{s-\delta} \psi_{s-\delta}^+ A_{s-\delta} X_{(s-\delta)-} \gamma_u^*]$ has negative components off the main diagonal, then so will $\Gamma_{s-\delta}^s$.

For $s$, a jump time, again for small $\delta > 0$

$$\Gamma_{s-\delta}^s = I + \alpha_s \psi_s^+ \Delta X_s \gamma_s^* + O(\delta).$$

The result follows.   $\square$

A useful result when applying this lemma is the following.

LEMMA 3.18.   *For a nonnegative column vector $x \in \mathbb{R}^K$ and any basis vector $e_i \in \mathbb{R}^K$, the matrix $I + x e_i^*$ has nonnegative components and is invertible.*

PROOF.   Clearly, $I + x e_i^*$ has nonnegative components. For $I + x e_i^*$ to be singular, there must be a nontrivial linear combination of its rows equalling zero. Suppose $y_j$ denotes the $j$th row of $I + x e_i^*$ and $\sum_{j=1}^K c_j y_j = \mathbf{0}$. As, for each $j \neq i$, $y_j$ is the only row containing a nonzero entry in the $j$th place, we must have $c_j = 0$ for $j \neq i$. Therefore we must have $c_i y_i = \mathbf{0}$. By assumption, $x$ is nonnegative, so $y_i = (1 + e_i^* x)e_i \neq \mathbf{0}$, therefore $c_i = 0$ also. Therefore there is no nontrivial linear combination of the rows of $I + x e_i^*$, and so $I + x e_i^*$ is invertible. $\quad \square$

LEMMA 3.19. *Under the conditions of Theorem 3.11,*

$$\sup_t E[\|\Gamma_0^t\|^2] < +\infty$$

*and*

$$\sup_t E\left[\left\|\Gamma_0^t Y_t + \int_{]0,t]} \Gamma_0^{u-} \phi_u \, du\right\|^2\right] < +\infty.$$

PROOF. We first show $\sup_t E\|\Gamma_0^t\|^2 < +\infty$. From (3.11),

$$E[\|\Gamma_0^t\|^2] \leq 2\|I\|^2 + 2E\left[\left\|\int_{]0,t]} \Gamma_0^{u-} [\beta_u \, du + \alpha_u \psi_u^+ \, dM_u \gamma_u^*]\right\|^2\right]$$

$$\leq 2\|I\|^2 + 4\int_{]0,t]} E[\|\Gamma_0^{u-}\|^2] E[\|\beta_u\|^2] \, du$$

$$+ 4E\left[\left\|\int_{]0,t]} \Gamma_0^{u-} \alpha_u \psi_u^+ \, dM_u \gamma_u^*\right\|^2\right].$$

The last term of the right-hand side can then be decomposed using Lemma 3.1, to give

$$4E\left[\left\|\int_{]0,t]} \Gamma_0^{u-} \alpha_u \psi_u^+ \, dM_u \gamma_u^*\right\|^2\right]$$

$$= 4E\left[\int_{]0,t]} (\gamma_u^* \gamma_u) \operatorname{Tr}(\Gamma_0^{u-} \alpha_u \psi_u^+ \, d\langle M, M\rangle_u (\psi_u^+)^* \alpha_u^* (\Gamma_0^{u-})^*)\right]$$

$$\leq 4E\left[\int_{]0,t]} (\gamma_u^* \gamma_u) \operatorname{Tr}((\Gamma_0^{u-})^* \Gamma_0^{u-}) \operatorname{Tr}(\alpha_u \psi_u^+ \psi_u (\psi_u^+)^* \alpha_u^*) \, du\right]$$

$$\leq 4\int_{]0,t]} E[\|\Gamma_0^{u-}\|^2] E[\|\gamma_u\|^2 \|\alpha_u \psi_u^+\|_{X_{u-}}^2] \, du.$$

We therefore have

$$E[\|\Gamma_0^t\|^2] \leq 2\|I\|^2 + 4\int_{]0,t]} E[\|\Gamma_0^{u-}\|^2] E[\|\beta_u\|^2 + \|\gamma_u\|^2 \|\alpha_u \psi_u^+\|_{X_{u-}}^2] \, du$$

and Grönwall's lemma yields

$$E[\|\Gamma_0^t\|^2] \leq 2\|I\|^2 \exp\left\{4E\left[\int_{]0,t]} \|\beta_u\|^2 + \|\gamma_u\|^2 \|\alpha_u \psi_u^+\|_{X_{u-}}^2 \, du\right]\right\}.$$

As $\alpha, \beta, \gamma, \psi$ and $\psi^+$ are all $du \times \mathbb{P}$-a.s. bounded (by assumption and Lemma 3.3), the right-hand side of this is finite for all $t$. As it is also increasing in $t$,

$$\sup_t E[\|\Gamma_0^t\|^2] \le 2\|I\|^2 \exp\left\{ 4E\left[ \int_{]0,T]} \|\beta_u\|^2 + \|\gamma_u\|^2 \|\alpha_u \psi_u^+\|_{X_{u-}}^2 \, du \right] \right\} < +\infty.$$

To show the second statement, note

$$E\left[ \left\| \Gamma_0^t Y_t + \int_{]0,t]} \Gamma_0^{u-} \phi_u \, du \right\|^2 \right]$$

$$\le 2E[\|\Gamma_0^t\|^2] E[\|Y_t\|^2] + 2\int_{]0,t]} E[\|\Gamma_0^{u-}\|^2] E[\|\phi_u\|^2] \, du.$$

Hence

$$\sup_t E\left[ \left\| \Gamma_0^t Y_t + \int_{]0,t]} \Gamma_0^{u-} \phi_u \, du \right\|^2 \right]$$

$$\le 2\left( \sup_t E[\|\Gamma_0^t\|^2] \right) \left( \sup_t E[\|Y_t\|^2] \right)$$

$$+ 2\left( \sup_u E[\|\Gamma_0^{u-}\|^2] \right) \left( E\left[ \int_{]0,T]} \|\phi_u\|^2 \, du \right] \right)$$

and each of these terms is finite. $\square$

PROOF OF THEOREM 3.11. It is clear that (3.8) is of the form of (1.3), where $F(\omega, t, Y_{t-}, Z_t) = \phi_t + \beta_t Y_{t-} + \alpha_t Z_t \gamma_t^*$ is a Lipschitz continuous, square integrable driver for the equation. The uniqueness of the solution follows from Theorem 1.1. We now search for the closed-form solution.

The first required result is that

$$\Gamma_0^t Y_t + \int_{]0,t]} \Gamma_0^u \phi_u \, du$$

is a uniformly integrable martingale. To see this, observe that in this context all integrals are Stieltjes integrals, and hence

$$\Gamma_0^t Y_t = \Gamma_0^0 Y_0 + \int_{]0,t]} [\Gamma_0^{u-} \, dY_u + d\Gamma_0^u Y_{u-}] + \sum_{0 < u \le t} \Delta \Gamma_0^u \Delta Y_u$$

$$= \Gamma_0^0 Y_0 - \int_{]0,t]} \Gamma_0^{u-} [\phi_u + \beta_u Y_{u-} + \alpha_u Z_u^* \gamma_u] \, du + \int_{]0,t]} \Gamma_0^{u-} Z_u \, dM_u$$

$$+ \int_{]0,t]} \Gamma_t^{u-} \beta_u Y_{u-} \, du + \int_{]0,t]} \Gamma_t^{u-} \alpha_u \psi_u^+ \, dM_u \gamma_u^* Y_{u-}$$

$$+ \sum_{0 < u \le t} \Gamma_0^{u-} \alpha_u \psi_u^+ \Delta M_u \gamma_u^* Z_u \Delta M_u.$$

Then, as a $1 \times 1$ matrix is its own transpose,

$$\Gamma_0^{u-} \alpha_u \psi_u^+ \Delta M_u \gamma_u^* Z_u \Delta M_u = \Gamma_0^{u-} \alpha_u \psi_u^+ \Delta M_u \Delta M_u^* Z_u^* \gamma_u.$$

The quantity $\Delta M_u \Delta M_u^*$ is equal to $d[M, M]_u$, the measure induced by the optional quadratic covariation matrix of $M$. $d\langle M, M \rangle_u$ is then the dual predictable projection of $d[M, M]_u$ (see [8]). Therefore,

$$\Gamma_0^t Y_t + \int_{]0,t]} \Gamma_0^{u-} \phi_u \, du = L_t - \int_{]0,t]} \Gamma_0^{u-} \alpha_u Z_u^* \gamma_u \, du$$
$$+ \int_{]0,t]} \Gamma_0^{u-} \alpha_u \psi_u^+ \, d\langle M, M \rangle_u Z_u^* \gamma_u$$

for some local martingale $L$. It follows that

$$\Gamma_0^t Y_t + \int_{]0,t]} \Gamma_0^{u-} \phi_u \, du = L_t + \int_{]0,t]} \Gamma_0^{u-} [\alpha_u \psi_u^+ \psi_u - \alpha_u] Z_u^* \gamma_u \, du$$

and from Lemmas 3.3 and 3.5, $\psi_u^+ \psi_u Z^* = Z^*$ without loss of generality, so the latter of these terms is zero, and, therefore, the left-hand side is a local martingale.

By Lemma 3.19, $\Gamma_0^t Y_t + \int_{]0,t]} \Gamma_0^u \phi_u \, du$ is square integrable, and hence is a uniformly integrable martingale (see [14], page 12). Hence it is indistinguishable from the conditional expectation of its terminal value (see [14], page 11)

$$\Gamma_0^t Y_t + \int_{]0,t]} \Gamma_0^{u-} \phi_u \, du = E\left[ \Gamma_0^T Q + \int_{]0,T]} \Gamma_0^u \phi_u \, du \Big| \mathcal{F}_t \right].$$

Through standard calculations and the use of Lemmas 3.12 and 3.15, we can conclude that, up to indistinguishability,

$$Y_t = E\left[ \Gamma_t^T Q + \int_{]t,T]} \Gamma_t^u \phi_u \, du \Big| \mathcal{F}_t \right]. \qquad \square$$

COROLLARY 3.20.  *If $Q$ and $\phi$ are nonnegative, and the assumptions of Theorem 3.16 are satisfied, then $Y$ is nonnegative. If, in addition $Y_0 = 0$, then, $Y = 0$ up to indistinguishability (and hence $Q = 0$, $\mathbb{P}$-a.s.) and $\phi_t = 0$, $\mathbb{P} \times dt$-a.s.*

PROOF.  This follows from Theorem 3.11 and the nonnegativity result of Theorem 3.16. The strict comparison is trivial, given the invertibility of $\Gamma_s^t$ for all $s$ and $t$ and the fact that $Y$ is càdlàg.  $\square$

**4. A scalar comparison theorem.** We shall now establish a comparison theorem relating the solutions of two BSDEs in the scalar case.

REMARK 4.1. In the scalar case, when $K = 1$, the assumptions of Theorem 3.16 simplify considerably. The assumption of nonnegativity of

$$\beta_t - \alpha_t \psi_t^+ A_t X_{t-} \gamma_t^*,$$

off the main diagonal becomes trivial (as there are no off diagonal terms) and the assumption that

$$I + \alpha_s \psi_s^+ (e_j - X_{s-}) \gamma_s^*$$

is $\mathbb{P}$-a.s. invertible and nonnegative for all $j$ such that $e_j^* A_s X_{s-} > 0$ can be simplified to

(4.1)                    $\alpha_s \psi_s^+ (e_j - X_{s-}) > -1$

as $\gamma = 1$, without loss of generality.

THEOREM 4.2 (Scalar comparison theorem). *Suppose we have two standard, scalar ($K = 1$), BSDEs corresponding to coefficients and terminal values $(F^1, Q^1)$ and $(F^2, Q^2)$. Let $(Y^1, Z^1)$ and $(Y^2, Z^2)$ be the associated solutions. We suppose the following conditions hold:*

 (i)  *$Q^1 \geq Q^2$ $\mathbb{P}$-a.s.;*
 (ii) *$dt \times \mathbb{P}$-a.s.,*

$$F^1(\omega, t, Y_{t-}^2, Z_t^2) \geq F^2(\omega, t, Y_{t-}^2, Z_t^2);$$

 (iii) *there exists an $\varepsilon > 0$ such that $\mathbb{P}$-a.s., for all $t \in [0, T]$, if $Z_t^1$, $Z_t^2$ are such that*

$$(e_j^* A_t X_{t-})[Z_t^1 - Z_t^2](e_j - X_{t-}) \geq -\varepsilon \| Z_t^1 - Z_t^2 \|_{X_{t-}}$$

*for all $e_j$, then*

$$F^1(\omega, t, Y_{t-}^2, Z_t^1) - F^1(\omega, t, Y_{t-}^2, Z_t^2) \geq -[Z_t^1 - Z_t^2] A_t X_{t-}$$

*with equality only if $\| Z_t^1 - Z_t^2 \|_{X_{t-}} = 0$.*

*It is then true that $Y^1 \geq Y^2$ $\mathbb{P}$-a.s. Moreover, this comparison is strict, that is, if on some $A \in \mathcal{F}_t$ we have $Y_t^1 = Y_t^2$, then $Q^1 = Q^2$ $\mathbb{P}$-a.s. on $A$, $F^1(\omega, s, Y_s^2, Z_s^2) = F^2(\omega, s, Y_s^2, Z_s^2)$ $ds \times \mathbb{P}$-a.s. on $[t, T] \times A$ and $Y^1$ is indistinguishable from $Y^2$ on $[t, T] \times A$.*

REMARK 4.3. Note that assumption (iii) need only hold for $Z^1$ and $Z^2$; there may well be other processes $Z$ for which it fails. This will correspond with various types of dominance, as will be seen in Section 7 below. In practice, however, it may be more convenient to assume that this assumption holds for all $Z^1$ and $Z^2$, as it is not clear how one might show this condition holds for a particular $(F, Q)$ without first finding the solution processes $(Y, Z)$ analytically.

An intuitive interpretation of this assumption is, "consider the difference between the SDEs with starting value $Y^2_{t-}$, trend $F(\omega, t, Y_t, Z^i_t)$, and hedging processes $Z^1_t$ and $Z^2_t$. If the only sizeable jumps that can occur in this difference are positive, then the overall trend through time should be negative."

PROOF OF THEOREM 4.2. We can write

$$
\begin{aligned}
(4.2) \quad & Y^1_t - Y^2_t - \int_{]t,T]} [F^1(\omega, u, Y^1_{u-}, Z^1_u) - F^2(\omega, u, Y^2_{u-}, Z^2_u)]\, du \\
& + \int_{]t,T]} [Z^1_u - Z^2_u]\, dM_u = Q_1 - Q_2.
\end{aligned}
$$

Taking (4.2), the equation satisfied by $\delta Y := Y^1 - Y^2$ and $\delta Z := Z^1 - Z^2$, we shall form an equivalent linear BSDE for $\delta Y$ and apply Corollary 3.20 to prove the desired result. For notational simplicity, we shall omit the $\omega, t$ arguments of $F^1$, $F^2$ as implicit. We also define $0/0 := 0$ wherever needed.

Without loss of generality, we can assume

$$
\varepsilon < \varepsilon_r^{3/2} N^{-3/2},
$$

where $\varepsilon_r$ is as in (3.1).

We consider three cases.

1. If $F^1(Y^2_{t-}, Z^1_t) - F^1(Y^2_{t-}, Z^2_t) \geq 0$, then let

$$
\begin{aligned}
\phi_t &= F^1(Y^2_{t-}, Z^1_t) - F^2(Y^2_{t-}, Z^2_t), \\
\beta_t &= \frac{F^1(Y^1_{t-}, Z^1_t) - F^1(Y^2_{t-}, Z^1_t)}{\delta Y_{t-}}, \\
\alpha_t &= 0, \\
\gamma_t &= 1.
\end{aligned}
$$

Note that assumption (ii) and the fact $F^1(Y^2_{t-}, Z^1_t) - F^1(Y^2_{t-}, Z^2_t) \geq 0$ implies that $\phi_t \geq 0$.

2. If $F^1(Y^2_{t-}, Z^1_t) - F^1(Y^2_{t-}, Z^2_t) < 0$ and there is a $j$ such that

$$
(e^*_j A_t X_{t-})[Z^1_t - Z^2_t](e_j - X_{t-}) < -\varepsilon \|Z^1_t - Z^2_t\|_{X_{t-}},
$$

then let

$$
\phi_t = F^1(Y^2_{t-}, Z^2_t) - F^2(Y^2_{t-}, Z^2_t),
$$

$$\beta_t = \frac{F^1(Y^1_{t-}, Z^1_t) - F^1(Y^2_{t-}, Z^1_t)}{\delta Y_{t-}},$$

$$\alpha_t = \frac{F^1(Y^2_{t-}, Z^1_t) - F^1(Y^2_{t-}, Z^2_t)}{[Z^1_t - Z^2_t]\psi_t e_j} \cdot e_j^* \psi_t,$$

$$\gamma_t = 1.$$

3. If $F^1(Y^2_{t-}, Z^1_t) - F^1(Y^2_{t-}, Z^2_t) < 0$ and

(4.3) $$(e_j^* A_t X_{t-})[Z^1_t - Z^2_t](e_j - X_{t-}) \geq -\varepsilon \|Z^1_t - Z^2_t\|_{X_{t-}}$$

for all $j$, then let

$$\phi_t = F^1(Y^2_{t-}, Z^2_t) - F^2(Y^2_{t-}, Z^2_t),$$

$$\beta_t = \frac{F^1(Y^1_{t-}, Z^1_t) - F^1(Y^2_{t-}, Z^1_t)}{\delta Y_{t-}},$$

$$\alpha_t = \frac{F^1(Y^2_{t-}, Z^1_t) - F^1(Y^2_{t-}, Z^2_t)}{[Z^1_t - Z^2_t]\psi_t X_{t-}} \cdot X_{t-}^* \psi_t,$$

$$\gamma_t = 1.$$

In all three cases, it is clear that

$$F^1(Y^1_{t-}, Z^1_t) - F^2(Y^2_{t-}, Z^2_t) = \phi_t + \beta_t(\delta Y_{t-}) + \alpha_t(\delta Z_t)^* \gamma_t,$$

and so the linear BSDE with these values of $\phi_t$, $\beta_t$, $\alpha_t$ and $\gamma_t$ is equivalent to (4.2).

Furthermore, $E[\int_{[0,T]} \|\phi_t\|^2 \, dt] < +\infty$ as $F$ is standard, and in each case, $\beta_t, \alpha_t$ and $\gamma_t$ are $dt \times \mathbb{P}$-a.s. bounded. This is trivial for $\gamma_t$ and follows directly from Lipschitz continuity for $\beta_t$ in all cases.

In case 1, $\alpha_t = 0$. In case 2, we know that $\psi_t e_j = (e_j^* A_t X_{t-})[e_j - X_{t-}]$ by the definition of $\psi$ in (1.4). By assumption, the absolute value of $[Z^1_t - Z^2_t]\psi_t e_j$ is then at least $\varepsilon \|Z^1_t - Z^2_t\|_{X_{t-}}$, and therefore $\alpha_t$ is bounded by Lipschitz continuity.

In case 3, by Lemma 3.6 and the fact $F^1(Y^2_{t-}, Z^1_t) \neq F^1(Y^2_{t-}, Z^2_t)$, we know $\|Z^1_t - Z^2_t\|_{X_{t-}} \neq 0$, and so we are in precisely the situation considered in Lemma 3.10. Therefore $[Z^1_t - Z^2_t]\psi_t X_{t-} < -\varepsilon \|Z^1_t - Z^2_t\|_{X_{t-}}$, and so Lipschitz continuity implies the boundedness of $\alpha_t$ in case 3.

We have assumed that $F^1$ and $F^2$ are standard, and hence $\mathbb{P}$-a.s. is left continuous in $t$, as are $Y^i_{t-}$ and $Z^i_t$ for $i = 1, 2$. By construction, we see that the variable $C_t$ indicating which of cases 1, 2 and 3 is in force at any time $t$ is predictable. Furthermore, the processes $\alpha$, $\beta$ and $\phi$ defined in each case are also predictable. We then use $C_t$ to piece together the various definitions of $\alpha$, $\beta$ and $\phi$ to give a single predictable linear BSDE, with the same driver values as $F^1(Y^1_{t-}, Z^1_t) - F^2(Y^2_{t-}, Z^2_t)$, $dt \times \mathbb{P}$-a.s. This linear BSDE satisfies all the requirements for Theorem 3.11.

We now appeal to Remark 4.1 to determine that the only requirement to show nonnegativity of $\delta Y$ is that, for all $t$, $\alpha_t \psi^+_t(e_k - X_{t-}) > -1$ for all $k$ with $e^*_k A_{t-} X_{t-} > 0$.

In case 1 this is clear. In case 2, we can write

$$\alpha_t \psi^+_t(e_k - X_{t-}) = \frac{F^1(Y^2_{t-}, Z^1_t) - F^1(Y^2_{t-}, Z^2_t)}{e^*_j \psi_t [Z^1_t - Z^2_t]^*} e^*_j \psi_t \psi^+_t(e_k - X_{t-}),$$

and, by Lemma 3.4, $e^*_j \psi_t \psi^+(e_k - X_{t-}) = e^*_j(e_k - X_{t-})$ which is zero unless $k = j$. If $k = j$ we see that

$$\alpha_t \psi^+_t(e_k - X_{t-}) = \frac{F^1(Y^2_{t-}, Z^1_t) - F^1(Y^2_{t-}, Z^2_t)}{e^*_j \psi_t [Z^1_t - Z^2_t]^*} > 0$$

by construction.

In case 3, we know from Lemma 3.10 that $[Z^1_t - Z^2_t]\psi_t X_{t-}$ is negative. For any $k$ satisfying Lemma 3.4, this implies

$$\frac{X^*_{t-}\psi_t \psi^+_t(e_k - X_{t-})}{X^*_{t-}\psi_t [Z^1_t - Z^2_t]^*} = \frac{X^*_{t-}(e_k - X_{t-})}{X^*_{t-}\psi_t [Z^1_t - Z^2_t]^*}$$

is positive. We then use assumption (iii) of the theorem, along with Lemma 3.7, to show that

$$\begin{aligned}
\alpha_t &\psi^+_t(e_k - X_{t-}) \\
&= \frac{F^1(\omega, t, Y^2_{t-}, Z^1_t) - F^1(\omega, t, Y^2_{t-}, Z^2_t)}{[Z^1_t - Z^2_t]\psi_t X_{t-}} X^*_{t-}(e_k - X_{t-}) \\
&> -\frac{[Z^1_t - Z^2_t]A_t X_{t-}}{[Z^1_t - Z^2_t]\psi_t X_{t-}} X^*_{t-}(e_k - X_{t-}) \\
&= -\frac{[Z^1_t - Z^2_t]\psi_t X_{t-}}{[Z^1_t - Z^2_t]\psi_t X_{t-}} \\
&= -1
\end{aligned}$$

as desired.

Therefore, we have shown that $\delta Y = Y^1 - Y^2$; the difference of our processes satisfies the requirements of Remark 4.1. That is, the assumptions of Corollary 3.20 are satisfied by this process and $\delta Y$ is therefore nonnegative $\mathbb{P}$-a.s. The rest of the theorem also follows by Corollary 3.20. $\square$

REMARK 4.4. In general assumption (iii) cannot be omitted, and is closely related to various geometric interpretations of no arbitrage (see Section 7). The fact that it is possible to create such $Z^1$ and $Z^2$ (due to the linear redundancy in $M$), indicates a significant difference between the Markov chain theory considered here and that based on Brownian motion considered elsewhere.

4.1. *Scalar counterexamples.* Suppose the theorem were to hold without assumption (iii). Then we can create contradictions as follows.

EXAMPLE 4.1. Take any standard $F$ such that there is $Y^1$, $Y^2$, $Z^1$ and $Z^2$ with $Y^1_0 = Y^2_0$, $[Z^1_t - Z^2_t](e_j - X_{t-})$ nonnegative for all $e_j$ with $e_j^* A_t X_{t-} > 0$ and positive for at least one such $j$, and

$$F(\omega, t, Y^2_{t-}, Z^1_t) - F(\omega, t, Y^2_{t-}, Z^2_t) \leq -[Z^1_t - Z^2_t]A_t X_{t-}$$

on some (stochastic) set of positive $dt$ measure. Without loss of generality, we shall assume that this set is the stochastic interval $[0, S]$, where $S$ is an essentially bounded stopping time. Extend $F$ to $[0, T]$ for some deterministic time $T$ as in Section 2.

Then define $Q^1$ and $Q^2$ by the forward stochastic differential equations

$$Y^1_t - \int_{]t,T]} F(\omega, u, Y^1_{u-}, Z^1_u)\, du + \int_{]t,T]} Z^1_u\, dM_u = Q^1,$$

$$Y^2_t - \int_{]t,T]} F(\omega, u, Y^2_{u-}, Z^2_u)\, du + \int_{]t,T]} Z^2_u\, dM_u = Q^2.$$

If we consider the process $\delta Y := Y^1 - Y^2$, we can see that $\delta Y$ is nondecreasing over $[0, T]$ with probability one, and is increasing with positive probability. This is because at every jump time $s$,

$$\Delta(\delta Y)_s = [Z^1_s - Z^2_s]\Delta X_s = [Z^1_s - Z^2_s](e_j - X_{s-})$$

for some $j$ which is nonnegative a.s. and positive with positive probability, and at every other time, the derivative of $\delta Y$ is equal to

$$\begin{aligned}
\frac{d(\delta Y)}{du} &= -F(\omega, u, Y^1_{u-}, Z^1_u) + F(\omega, u, Y^2_{u-}, Z^2_u)\\
&\quad - Z^1_u A_u X_{u-} + Z^2_u A_u X_{u-}\\
&> 0
\end{aligned}$$

by assumption. Therefore $Q^1 \geq Q^2$ $\mathbb{P}$-a.s., and $Q^1 > Q^2$ with positive probability. However, the initial value, $\delta Y_0 = 0$, would imply, by the strict comparison of the theorem, that $Q^1 = Q^2$ $\mathbb{P}$-a.s., which is a contradiction.

EXAMPLE 4.2. For a given Markov chain, $X$, with three states where the rate matrix is such that the jump rate between any pair of states is nonvanishing, define the process $\{J_t\}$ which counts the number of jumps in $X$ strictly prior to time $t$. Let

$$Z^1_t = [2\sqrt{1 - 2^{-2(J_t+2)}}e^*_{i_1} - 2^{-(J_t+1)}e^*_{i_2}]\psi_t \psi^+_t,$$

where $e_{i_1}$, $e_{i_2}$ and $X_{t-}$ are the three basis vectors in $\mathbb{R}^3$. Let $Z^2 = 0$ and

$$F^1 = F^2 = F(\omega, t, Y_{t-}, Z_t) = -\|Z_t\|_{X_{t-}} - Z_t A_t X_{t-}.$$

This particular combination for $Z^1 - Z^2$ always has the possibility of negative jumps, however, these decrease exponentially in magnitude while $\|Z^1 - Z^2\|_{X_{t-}} = 4$ is constant. For $J_t$ sufficiently large, this will fail to satisfy assumption (iii) for any fixed $\varepsilon$. The sizes of the negative jumps are such that their sum will always be more than $-1$.

Consider the forward SDEs defined by these $Z^1$ and $Z^2$ with starting value $Y_0^1 = Y_0^2 = 0$; for terminal time $T = 1$, $Y_T^2 = 0$ and $Y_T^1 \geq 3$ $\mathbb{P}$-a.s., contradicting the theorem.

REMARK 4.5. These examples also demonstrate the interrelationship between the scalar comparison theorem, assumption (iii) and dominance. This is explored further in Section 7.

**5. General comparison theorems.** We now wish to extend Theorem 4.2 to the vector case. This is nontrivial, as the simplifications of Remark 4.1 are not possible, and we must satisfy the more difficult conditions of Corollary 3.20 directly. We also do not have the useful result of Lemma 3.10. We present here some alternative generalizations.

THEOREM 5.1 (Vector comparison Theorem 1). *Suppose we have two standard BSDE parameters, $(F^1, Q^1)$ and $(F^2, Q^2)$. Let $(Y^1, Z^1)$ and $(Y^2, Z^2)$ be the associated solutions.*

*Recall that a BSDE here has $F(\omega, t, Y_{t-}, Z_t), Y_{t-}, Q \in \mathbb{R}^K$ and $Z_t \in \mathbb{R}^{K \times N}$ where $N$ is the number of states of the Markov chain $X$.*

*We suppose the following conditions hold:*

(i) $Q^1 \geq Q^2$ $\mathbb{P}$-a.s.;

(ii) $dt \times \mathbb{P}$-a.s.,

$$F^1(\omega, t, Y_{t-}^2, Z_t^2) \geq F^2(\omega, t, Y_{t-}^2, Z_t^2);$$

(iii) *there exists an $\varepsilon > 0$ such that $\mathbb{P}$-a.s., for all $t \in [0, T]$, for any basis vector $e_k \in \mathbb{R}^K$, if $Z_t^1$, $Z_t^2$ are such that*

$$(e_j^* A_t X_{t-}) e_k^* [Z_t^1 - Z_t^2](e_j - X_{t-}) \geq -\varepsilon \|e_k^* [Z_t^1 - Z_t^2]\|_{X_{t-}}$$

*for all $e_j$, then*

$$e_k^* [F^1(\omega, t, Y_{t-}^2, Z_t^1) - F^1(\omega, t, Y_{t-}^2, Z_t^2)] \geq -e_k^* [Z_t^1 - Z_t^2] A_t X_{t-}$$

*with equality only if $\|e_k^* [Z_t^1 - Z_t^2]\|_{X_{t-}} = 0$;*

(iv) *for each $i$, $e_i^* F^1$ can be written as a function*

$$e_i^* F^1(\omega, t, Y_{t-}, Z_t) = F_i^1(\omega, t, e_i^* Y_{t-}, e_i^* Z_t),$$

*that is, the $i$th component of $F^1$ depends only on the $i$th component of $Y$ and the $i$th row of $Z$ (and $\omega$ and $t$); and similarly for $F^2$.*

It is then true that $Y^1 \geq Y^2$ $\mathbb{P}$-a.s. Moreover, this comparison is strict, that is, if on some $A \in \mathcal{F}_t$, we have $Y_t^1 = Y_t^2$, then $Q^1 = Q^2$ $\mathbb{P}$-a.s. on $A$, $F^1(\omega, s, Y_s^2, Z_s^2) = F^2(\omega, s, Y_s^2, Z_s^2)$ $ds \times \mathbb{P}$-a.s. on $[t, T] \times A$ and and $Y^1$ is indistinguishable from $Y^2$ on $[t, T] \times A$.

PROOF. This can be established using the results of Theorem 4.2. From (1.3), we know that, with $\delta Y := Y^1 - Y^2$, $\delta Z := Z^1 - Z^2$,

$$Q^1 - Q^2 = \delta Z_t - \int_{]t,T]} [F^1(\omega, u, Z_{u-}^1, Y_u^1) - F^2(\omega, u, Z_{u-}^2, Y_u^2)] \, du$$

$$+ \int_{]t,T]} \delta Y_u \, dM_u,$$

which implies, for each $i$,

$$e_i^*[Q^1 - Q^2]$$

$$= e_i^* \delta Y_t - \int_{]t,T]} e_i^* [F^1(\omega, u, Y_{u-}^1, Z_u^1) - F^2(\omega, u, Y_{u-}^2, Z_u^2)] \, du$$

$$+ \int_{]t,T]} e_i^* \delta Z_u \, dM_u$$

$$= e_i^* \delta Y_t - \int_{]t,T]} [F_i^1(\omega, u, e_i^* Y_{u-}^1, e_i^* Z_u^1) - F_i^2(\omega, u, e_i^* Y_{u-}^2, e_i^* Z_u^2)] \, du$$

$$+ \int_{]t,T]} e_i^* \delta Z_u \, dM_u,$$

which is of the form considered in Theorem 4.2. As all the assumptions needed in the scalar case are given, the result is shown for each component $e_i^* \delta Y$ by Theorem 4.2. As this holds for all $i$, the vector result follows. $\square$

COROLLARY 5.2. *For any pair of BSDEs satisfying assumption (iv) of Theorem 5.1, if assumptions (i)–(iii) are satisfied by any component of the terminal condition and driver, then the comparison theorem will hold on that component.*

PROOF. This is a direct result of applying Theorem 4.2 to the component in question. $\square$

It is clear that this theorem does not provide for much more behavior than the scalar case, as it is considering each component separately from the others. Removing this requirement is nontrivial, but can be done. To do so using the machinery of the solutions to linear BSDEs, we must make the following, in some sense stronger, assumption.

THEOREM 5.3 (Vector comparison Theorem 2). *Suppose we have two standard BSDE parameters $(F^1, Q^1)$ and $(F^2, Q^2)$. Let $(Y^1, Z^1)$ and $(Y^2, Z^2)$ be the associated solutions where the dimensions of $F$, $Q$, $Y$ and $Z$ are as in Theorem 5.1. We suppose the following conditions hold:*

(i) *$Q^1 \geq Q^2$ $\mathbb{P}$-a.s.;*

(ii) *$dt \times \mathbb{P}$-a.s.,*

$$F^1(\omega, t, Y_{t-}^2, Z_t^2) \geq F^2(\omega, t, Y_{t-}^2, Z_t^2);$$

(iii) *there exists an $\varepsilon > 0$ such that, $\mathbb{P}$-a.s., for all $t \in [0, T]$, for any basis vector $e_j \in \mathbb{R}^K$, if*

$$(e_k^* A_t X_{t-}) e_j^* [Z_t^1 - Z_t^2](e_k - X_{t-}) \geq -\varepsilon \|Z_t^1 - Z_t^2\|_{X_{t-}}$$

*for all $e_k \neq X_{t-}$, then*

$$e_j^* [F^1(\omega, t, Y_{t-}^2, Z_t^1) - F^1(\omega, t, Y_{t-}^2, Z_t^2)] \geq 0;$$

(iv) *for each $i$, $e_i^* F^1$ can be written as a function*

$$e_i^* F^1(\omega, t, Y_{t-}, Z_t) = F_i^1(\omega, t, e_i^* Y_{t-}, Z_t),$$

*that is, the $i$th component of $F^1$ depends only on the $i$th component of $Y$ (and on $\omega, t$ and $Z_t$).*

*It is then true that $Y^1 \geq Y^2$ $\mathbb{P}$-a.s. Moreover, this comparison is strict, that is, if on some $A \in \mathcal{F}_t$ we have $Y_t^1 = Y_t^2$, then $Q^1 = Q^2$ $\mathbb{P}$-a.s. on $A$, $F^1(\omega, s, Y_s^2, Z_s^2) = F^2(\omega, s, Y_s^2, Z_s^2)$ $ds \times \mathbb{P}$-a.s. on $[t, T] \times A$, and $Y^1$ is indistinguishable from $Y^2$ on $[t, T] \times A$.*

REMARK 5.4. The key differences between Theorems 5.3 and 5.1 are that in Theorem 5.3, a weaker assumption is placed on the behavior of the driver $F^1$ in relation to interactions between the rows of the $Z$ matrix, but a stronger assumption is placed on the behavior of a component of $F^1$ when no jump is significantly negative in that component. We shall also see that this weaker assumption leads to a significantly weaker result—the strict comparison cannot here be shown to hold componentwise.

REMARK 5.5.   Assumption (iii) in Theorem 5.3 is a stronger assumption than for Theorem 5.1. In the scalar case, which is the foundation of Theorem 5.1, we used Lemma 3.10 to show that each row, $e_i^*[Z_t^1 - Z_t^2]\psi_t X_{t-}$, is negative, and our assumption was equivalent to

$$e_k^*[F^1(\omega, t, Y_{t-}^2, Z_t^1) - F^1(\omega, t, Y_{t-}^2, Z_t^2)] \geq e_k^*[Z_t^1 - Z_t^2]\psi_t X_{t-},$$

which is clearly satisfied when

$$e_k^*[F^1(\omega, t, Y_{t-}^2, Z_t^1) - F^1(\omega, t, Y_{t-}^2, Z_t^2)] \geq 0.$$

PROOF OF THEOREM 5.3.   For simplicity of notation, again let $\delta Y_t = Y_t^1 - Y_t^2$, $\delta Z_t = Z_t^1 - Z_t^2$. We also will omit the $\omega$ and $t$ arguments of $F$ as implicit. Note that here $\phi, \gamma \in \mathbb{R}^K$, $\beta \in \mathbb{R}^{K \times K}$ and $\alpha \in \mathbb{R}^{K \times N}$. We will construct $\alpha$, $\beta$, $\gamma$ and $\phi$ in two stages.

1. Consider first those components of $F^1(Y_{t-}^2, Z_t^1) - F^1(Y_{t-}^2, Z_t^2)$ which are nonnegative. Let $e_i$ indicate each such component in turn. Then define the corresponding components of $\phi_t$ and $\beta_t$, and rows of $\alpha_t$, by

$$e_i^* \phi_t = e_i^*[F^1(Y_{t-}^2, Z_t^1) - F^2(Y_{t-}^2, Z_t^2)] \geq 0,$$

$$e_i^* \beta_t = \frac{e_i^*[F^1(Y_{t-}^1, Z_t^1) - F^1(Y_{t-}^2, Z_t^1)]}{e_i^*[\delta Y_{t-}]} e_i^*,$$

$$e_i^* \alpha_t = 0.$$

It follows that $E[\int_{]0,T]} \|e_i^* \phi_t\|^2 \, dt] < +\infty$ as $F^1, F^2$ are standard, $e_i^* \beta_t$ is $dt \times \mathbb{P}$-a.s. bounded by Lipschitz continuity of $F^1$, and they will satisfy the requirements of Corollary 3.20.

2. If all components are dealt with by the above, then we have completed the construction. Otherwise, for each time point $t$, let $\gamma_t$ indicate the smallest of the remaining components of $\delta Y_{t-}$ in absolute value, that is, $\gamma_t$ is the basis vector such that $|\gamma_t^* \delta Y_{t-}| \leq |e_j^* \delta Y_{t-}|$ for all $j$.

For each of the remaining components, we now define

$$e_i^* \phi_t = e_i^*[F^1(Y_{t-}^2, Z_t^2) - F^2(Y_{t-}^2, Z_t^2)] \geq 0,$$

$$e_i^* \beta_t = \frac{e_i^*[F^1(Y_{t-}^1, Z_t^1) - F^1(Y_{t-}^2, Z_t^1)] - B\gamma_t^*[\delta Y_{t-}]}{e_i^*[\delta Y_{t-}]} e_i^* + B\gamma_t^*$$

for a large, fixed number $B \geq c\varepsilon^{-1}\varepsilon_r^{-1}$, where $c$ is the Lipschitz constant of $F^1$, $\varepsilon$ is as in assumption (iii) in the theorem, and $\varepsilon_r$ is as in (3.1). As $\gamma_t$ indicates the smallest component in absolute value of $Y_{t-}$, Lipschitz continuity again guarantees that this will remain bounded.

By assumption (iii) of the theorem, recalling

$$\psi_t e_k = (e_k A_t X_{t-})[e_k - X_{t-}],$$

either for some $e_k \neq X_{t-}$,

(5.1) $$\gamma_t^*[\delta Z_t]\psi_t e_k < -\varepsilon\|\delta Z_t\|_{X_{t-}}$$

or

$$\gamma_t^*[F^1(Y_{t-}^2, Z_t^1) - F^1(Y_{t-}^2, Z_t^2)] \geq 0.$$

However, we have selected $\gamma_t$ such that

(5.2) $$\gamma_t^*[F^1(Y_{t-}^2, Z_t^1) - F^1(Y_{t-}^2, Z_t^2)] < 0$$

and therefore (5.1) must hold. Define

$$e_i^*\alpha_t = \left(\frac{e_i^*[F^1(Y_{t-}^2, Z_t^1) - F^1(Y_{t-}^2, Z_t^2)]}{\gamma_t^*[\delta Z_t]\psi_t e_k}\right)e_k^*\psi_t.$$

This is $dt \times \mathbb{P}$-a.s. bounded by Lipschitz continuity and (5.1). As $e_k^*\psi_t\psi_t^+(e_j - X_{t-}) \geq 0$ for $e_j, e_k \neq X_{t-}$, and the fraction in parentheses is positive [by (5.1)] we have that

$$I + \alpha_k\psi^+(e_j - X_{t-})\gamma^*$$

is an invertible matrix with nonnegative components by Lemmas 3.18 and 3.4.

For the rows defined in the first case, we have $\beta_t$ only having elements on the main diagonal, and $\alpha_t$ having a corresponding row of zeros. Hence $\beta_t - \alpha_t\psi_t^+ A_t X_{t-}\gamma_t^*$ has only entries on the main diagonal. For the rows defined in the second case, we have, for each row $i$,

$$|e_i^*\alpha_t\psi_t^+ A_t X_{t-}| = \left|\left(\frac{e_i^*[F^1(Y_{t-}^2, Z_t^1) - F^1(Y_{t-}^2, Z_t^2)]}{\gamma_t^*[\delta Z_t]\psi_t e_k}\right)e_k^*\psi_t\psi_t^+ A_t X_{t-}\right|$$

$$\leq \left|\frac{e_i^*[F^1(Y_{t-}^2, Z_t^1) - F^1(Y_{t-}^2, Z_t^2)]}{\gamma_t^*[\delta Z_t]\psi_t e_k}\right|\varepsilon_r^{-1}$$

$$\leq c\varepsilon^{-1}\varepsilon_r^{-1}$$

$$\leq B.$$

Therefore the $i$th row of $\beta_t - \alpha_t\psi_t^+ A_t X_{t-}\gamma_t^*$ has an entry on the main diagonal, an entry of $B - e_i^*\alpha_t\psi_t^+ A_t X_{t-} \geq 0$ in the $\gamma_t$ column (when this is off the main diagonal), and zeros elsewhere. Hence $\beta_t - \alpha_t\psi_t^+ A_t X_{t-}\gamma_t^*$ has nonnegative entries off the main diagonal.

We therefore have defined, for each time $t$, random variables $\phi_t$, $\beta_t$, $\alpha_t$ such that

$$F^1(Y_{t-}^1, Z_t^1) - F^2(Y_{t-}^2, Z_t^2) = \phi_t + \beta_t\delta Y_{t-} + \alpha_t\delta Z_t^*\gamma_t.$$

As in the scalar case, we can now piece together these $\phi_t$, $\beta_t$, $\alpha_t$ and $\gamma_t$ to give predictable processes, $\phi$, $\beta$, $\alpha$ and $\gamma$. We have established that $\beta$, $\alpha$ and $\gamma$ are $dt \times \mathbb{P}$-a.s. bounded. As $F$ is standard, $\phi$ is also square integrable, and is nonnegative by construction.

Therefore, up to the desired $dt \times \mathbb{P}$-a.s. level of certainty, the requirements of Corollary 3.20 are satisfied for the linear BSDE solved by $\delta Y$. The result follows. □

REMARK 5.6. This proof is significantly weaker than that of Theorem 5.1. In particular, Corollary 5.2 has not been established in this case. It would appear intuitively reasonable, from a geometric perspective, that it should hold, and also that the weaker conditions of Theorem 5.1 assumption (iii) should be sufficient. However, the machinery of appealing to the solutions of linear BSDEs does not appear to be adequate to prove these results.

The assumption that the $i$th component of $F$ can depend only on the $i$th component of $Y$ may be overly restrictive. Because of this, we have the following alternative generalization, where we instead assume that $F$ does not depend on $Z$.

THEOREM 5.7 (Vector comparison Theorem 3). *Suppose we have two standard BSDE parameters $(F^1, Q^1)$ and $(F^2, Q^2)$. Let $(Y^1, Z^1)$ and $(Y^2, Z^2)$ be the associated solutions. We suppose the following conditions hold:*

(i) *$Q^1 \geq Q^2$ $\mathbb{P}$-a.s.;*

(ii) *$dt \times \mathbb{P}$-a.s.,*

$$F^1(\omega, t, Y_{t-}^2, Z_t^2) \geq F^2(\omega, t, Y_{t-}^2, Z_t^2);$$

(iii) *there exists an $\varepsilon > 0$ such that, $dt \times \mathbb{P}$-a.s., for each $i$, if*

$$e_i^* F^1(\omega, t, Y_{t-}^1, Z_t) < e_i^* F^1(\omega, t, Y_{t-}^2, Z_t),$$

*then either*

$$|e_i^*[Y_{t-}^1 - Y_{t-}^2]| > \varepsilon \|Y_{t-}^1 - Y_{t-}^2\|$$

*or there is a $j$ with*

$$e_j^*[Y_{t-}^1 - Y_{t-}^2] < -\varepsilon \|Y_{t-}^1 - Y_{t-}^2\|;$$

(iv) *$F^1$ does not depend on $Z$ (or equivalently, by Lemma 3.9, $F^1$ depends only on the row sums of $Z$, not on its individual elements).*

*It is then true that $Y^1 \geq Y^2$ $\mathbb{P}$-a.s. Moreover, this comparison is strict, that is, if on some $A \in \mathcal{F}_t$ we have $Y_t^1 = Y_t^2$, then $Q^1 = Q^2$ $\mathbb{P}$-a.s. on $A$, $F^1(\omega, s, Y_s^2, Z_s^2) = F^2(\omega, s, Y_s^2, Z_s^2)$ $ds \times \mathbb{P}$-a.s. on $[t, T] \times A$ and $Y^1$ is indistinguishable from $Y^2$ on $[t, T] \times A$.*

PROOF.   We have that

$$
\begin{aligned}
(5.3) \quad & Y_t^1 - Y_t^2 - \int_{]t,T]} [F^1(\omega, u, Y_{u-}^1, Z_u^1) - F^2(\omega, u, Y_{u-}^2, Z_u^2)]\, du \\
& + \int_{]t,T]} [Z_u^1 - Z_u^2]\, dM_u = Q_1 - Q_2.
\end{aligned}
$$

As before, let $\delta Y = Y^1 - Y^2$, and omit the $\omega$ and $t$ arguments of $F$ as implicit.

We first seek to construct an $\mathbb{R}^K$ process $\phi_t$ and an $\mathbb{R}^{K \times K}$ process $\beta_t$ such that the linear BSDE with these components matches (5.3). As $F^1$ does not depend on $Z$, we shall see that the $\alpha$ term in this BSDE is zero. Again, consider first those components, represented by $e_i$, where

$$
e_i^*[F^1(Y_{t-}^1, Z_t^1) - F^1(Y_{t-}^2, Z_t^1)] \geq 0.
$$

For these rows, define

$$
\begin{aligned}
(5.4) \quad & e_i^* \phi_t = e_i^*[F^1(Y_{t-}^1, Z_t^1) - F^2(Y_{t-}^2, Z_t^2)] \geq 0, \\
& e_i^* \beta_t = \mathbf{0}.
\end{aligned}
$$

For the other rows, first define

$$
e_i^* \phi_t = e_i[F^1(Y_{t-}^2, Z_t^2) - F^2(Y_{t-}^2, Z_t^2)] \geq 0.
$$

As we are in the situation considered in assumption (iii), either

$$
|e_i^*[Y_{t-}^1 - Y_{t-}^2]| > \varepsilon \|Y_{t-}^1 - Y_{t-}^2\|,
$$

in which case define

$$
(5.5) \quad e_i^* \beta_t = \frac{e_i^*[F^1(Y_{t-}^1, Z_t^2) - F^1(Y_{t-}^2, Z_t^2)]}{e_i^*[\delta Y_{t-}]} e_i^*
$$

or there is a $j$ with

$$
e_j^*[Y_{t-}^1 - Y_{t-}^2] < -\varepsilon \|Y_{t-}^1 - Y_{t-}^2\|,
$$

in which case, define

$$
(5.6) \quad e_i^* \beta_t = \frac{e_i^*[F^1(Y_{t-}^1, Z_t^2) - F^1(Y_{t-}^2, Z_t^2)]}{e_j^*[\delta Y_{t-}]} e_j^*.
$$

In both cases, $F$ is Lipschitz continuous, and so the matrix $\beta_t$ is bounded. We have constructed $\beta_t$ such that all off diagonal entries are are either zero [as in (5.4) and (5.5)], or are the ratio of two negative quantities [as in (5.6)], and will therefore be nonnegative.

We now note that, as assumption (iv) states, $F^1$ does not depend on $Z$, we have

$$F^1(Y^1_{t-}, Z^1_t) - F^1(Y^1_{t-}, Z^2_t) = F^1(Y^1_{t-}, 0) - F^1(Y^1_{t-}, 0)$$
$$= 0.$$

We again, as in Theorems 4.2 and 5.3, piece together these random variables $\phi_t$ and $\beta_t$ into a pair of predictable processes. As shown above, the process $\beta$ is bounded and has nonnegative quantities off the main diagonal, the process $\phi$ is nonnegative by construction and $E[\int_{]0,T]} \|\phi_t\|^2 \, dt] < +\infty$ as $F$ is standard.

The process $\delta Y$ then satisfies the linear BSDE,

$$\delta Y_t - \int_{]t,T]} [\phi_t + \beta_t \delta Y_{u-}] \, du + \int_{]t,T]} [Z^1_u - Z^2_u] \, dM_u = Q_1 - Q_2 \geq 0.$$

As $\beta_t$ has nonnegative entries off the main diagonal and $\alpha_t \equiv 0$, the conditions of Corollary 3.20 are satisfied. The result follows. $\square$

This theorem is counterintuitive, and when examined closely would appear to create a contradiction. The only resolution to this is the following corollary.

COROLLARY 5.8.  *For a pair of BSDEs satisfying Theorem 5.7, the strict comparison must hold componentwise. That is, if for some $t$ and some $e_i$, on some $A \in \mathcal{F}_t$ we have $e_i^* Y^1_t = e_i^* Y^2_t$, then $e_i^* Q^1 = e_i^* Q^2$ $\mathbb{P}$-a.s. on $A$, $e_i^* F^1(\omega, s, Y^2_s, Z^2_s) = e_i^* F^2(\omega, s, Y^2_s, Z^2_s) \, ds \times \mathbb{P}$-a.s. on $[t, T] \times A$ and $e_i^* Y^1$ is indistinguishable from $e_i^* Y^2$ on $[t, T] \times A$.*

PROOF.  Without loss of generality, we shall assume that $i = 1$, $t = 0$, $A = \Omega$. Then the assumptions of the theorem state $e_1^* Y^1_0 = e_1^* Y^2_0$, and from Theorem 5.7, $Y^1_s \geq Y^2_s$ $\mathbb{P}$-a.s. for all $s \geq 0$. These assumptions immediately imply that, for small $s$, the process $e_1^* Y_s$ must be $\mathbb{P}$-a.s. nondecreasing.

We can then rewrite (5.3) as a forward SDE for $\delta Y = Y^1 - Y^2$, with initial time $s_0$, omitting the $\omega$, $s$ and $Z$ arguments of $F$ as implicit or irrelevant,

$$(5.7) \quad \delta Y_s = \delta Y_{s_0} - \int_{]s_0,s]} [F^1(Y^1_{u-}) - F^2(Y^2_{u-})] \, du + \int_{]s_0,s]} [Z^1_u - Z^2_u] \, dM_u.$$

By assumption (iii) of Theorem 5.7, and the fact $Y^1_s \geq Y^2_s$, unless

$$|e_1^*[Y^1_{s-} - Y^2_{s-}]| > \varepsilon \|Y^1_{s-} - Y^2_{s-}\|$$

for some $\varepsilon > 0$, we must have

$$e_1^* F^1(Y^1_{s-}) \geq e_1^* F^1(Y^2_{s-})$$

$dt \times \mathbb{P}$-a.s. This implies that, $dt \times \mathbb{P}$-a.s.,

$$(5.8) \qquad -e_1^* F^1(Y_{s-}^1) + e_1^* F^2(Y_{s-}^2) \le -e_1^* F^1(Y_{s-}^2) + e_1^* F^2(Y_{s-}^2),$$

and by assumption (ii) of Theorem 5.7, this last term is nonpositive.

Now let $s^* = \inf\{s : e_1^* \delta Y_s \ne 0\} \wedge T$. Then as $\delta Y$ is càdlàg, $e_1^* \delta Y_{s^*-} = 0$. We wish to show $s^* = T$.

For $s = s^*$, the fact $\delta Y \ge 0$ and $e_1^* \delta Y_{s^*-} = 0$ implies, by assumption (iii) of Theorem 5.7, either that:

1. $$\delta Y_{s^*-} = 0$$

   or that

2. $$0 = |e_1^*[Y_{s^*-}^1 - Y_{s^*-}^2]| < \varepsilon \|Y_{s^*-}^1 - Y_{s^*-}^2\|$$

   for all $\varepsilon > 0$, and therefore (5.8) holds.

In the first case, we can use the strict comparison of Theorem 5.7 to show that $\delta Y_s = 0$ for all $s \ge s^*$. Therefore $s^* = T$.

In the second case, for simplicity of notation, for $s < 0$, we define, $Y_s^i = Y_0^i$, $Z_s^i = 0$, $F^i(\omega, s, Z, Y) = 0$ for $i = 1, 2$.

Suppose $s^* < T$, so $s^* = \inf\{s : e_1^* \delta Y_s \ne 0\}$. For small $\Delta s, \varepsilon > 0$, we can rewrite (5.7) as

$$e_1^* \delta Y_{s^*+\Delta s} = e_1^* \delta Y_{s^*-\varepsilon} - e_1^*[F^1(Y_{s^*-\varepsilon}^1) - F^2(Y_{s^*-\varepsilon}^2)](\Delta s + s^* - (s^* - \varepsilon) \vee 0)$$
$$+ O((\varepsilon + \Delta s)^2) + \int_{]s^*-\varepsilon, s^*+\Delta s]} e_1^*[Z_u^1 - Z_u^2] \, dM_u.$$

Note for $s^* > 0$ and $0 < \varepsilon < s^*$,

$$\Delta s + s^* - (s^* - \varepsilon) \vee 0 = \Delta s + \varepsilon.$$

We know from (5.8) that the term $-e_1^*[F^1(Y_{s^*-\varepsilon}^1) - F^2(Y_{s^*-\varepsilon}^2)](\Delta s + s^* - (s^* - \varepsilon) \vee 0)$ is nonincreasing in $\Delta s$, and therefore, as $\int_{]s^*-\varepsilon, s^*+\Delta s]} e_1^*[Z_u^1 - Z_u^2] \, dM_u$ is a martingale in $\Delta s$,

$$-e_1^*[F^1(Y_{s^*-\varepsilon}^1) - F^2(Y_{s^*-\varepsilon}^2)](\Delta s + s^* - (s^* - \varepsilon) \vee 0)$$
$$+ \int_{]s^*-\varepsilon, s^*+\Delta s]} e_1^*[Z_u^1 - Z_u^2] \, dM_u$$

cannot be $\mathbb{P}$-a.s. nondecreasing in $\Delta s$ except if it is zero.

However, this implies that for all sufficiently small $\Delta s$,

$$e_1^* \delta Y_{s^*+\Delta s} = e_1^* \delta Y_{s^*-\varepsilon} = 0.$$

This contradicts our assumption $s^* = \inf\{s : e_1^* \delta Y_s \ne 0\}$, and hence $s^* = T$ in the second case.

Therefore, we must have $e_1^* \delta Y = 0$ constant. In other words, we must have $e_1^*[Y^1 - Y^2] = 0$, that is, the strict comparison holds componentwise.  $\square$

5.1. *A vector example.*

EXAMPLE 5.1. To demonstrate the conclusion of Theorem 5.7, we consider the following example. Consider two BSDEs with the same driver,

$$F(\omega, t, Y_{t-}, Z_t) = f(\omega, t) + \begin{bmatrix} 0 & 0 \\ 1 & 1 \end{bmatrix} Y_{t-}$$

for a left continuous, $dt \times \mathbb{P}$ square integrable, $\mathbb{R}^2$ process $f = [f_1, f_2]^*$. Suppose that $Q^1 \geq Q^2$ $\mathbb{P}$-a.s. and that the terminal time $T$ is deterministic. Under these conditions, Theorem 5.7 should hold. Here the difference $Y^1 - Y^2$ satisfies the equation

$$Y_t^1 - Y_t^2 - \int_{]t,T]} \begin{bmatrix} 0 & 0 \\ 1 & 1 \end{bmatrix} (Y_{u-}^1 - Y_{u-}^2) \, du + \int_{]t,T]} [Z_u^1 - Z_u^2] \, dM_u = Q^1 - Q^2.$$

Our results on linear BSDEs, along with the fact that

$$\Pi_{]s,t]} \left\{ I + \begin{bmatrix} 0 & 0 \\ 1 & 1 \end{bmatrix} du \right\} = \exp \left\{ \begin{bmatrix} 0 & 0 \\ 1 & 1 \end{bmatrix} (t-s) \right\}$$

$$= I + (e^{t-s} - 1) \begin{bmatrix} 0 & 0 \\ 1 & 1 \end{bmatrix}$$

implies

$$Y_t^1 - Y_t^2 = E \left[ \left( I + (e^{T-t} - 1) \begin{bmatrix} 0 & 0 \\ 1 & 1 \end{bmatrix} \right) (Q^1 - Q^2) \Big| \mathcal{F}_t \right]$$

$$= \begin{bmatrix} 1 & 0 \\ e^{T-t} - 1 & e^{T-t} \end{bmatrix} E[Q^1 - Q^2 | \mathcal{F}_t].$$

It is then clear that the first component of $Y^1 - Y^2$ will be simply the conditional expectation. If this reaches zero for some $t$, as $Q^1 \geq Q^2$ $\mathbb{P}$-a.s., then the first component of $Q^1 - Q^2$ must be almost surely zero given $\mathcal{F}_t$. On the other hand, the second component is a positive sum of the two components of $Q^1 - Q^2$. If this equals zero for some $t$, then both components of $Q^1 - Q^2$ must be almost surely zero given $\mathcal{F}_t$.

Conversely, if we had

$$F(\omega, t, Y_{t-}, Z_t) = f(\omega, t) + \begin{bmatrix} 0 & 0 \\ -1 & 1 \end{bmatrix} Y_{t-}.$$

Theorem 5.7 would not apply. In this case we would find

$$Y_t^1 - Y_t^2 = \begin{bmatrix} 1 & 0 \\ -e^{T-t} + 1 & e^{T-t} \end{bmatrix} E[Q^1 - Q^2 | \mathcal{F}_t].$$

If then the second component of $Q^1 - Q^2$ has small conditional expectation compared with the first, for any $t < T$ we see that the second component of $Y^1 - Y^2$ is negative.

REMARK 5.9. A possible economic interpretation of the third assumption of Theorem 5.7 is the following. Suppose that $Y_t$ represents a vector of prices. If one asset has an increased price, that should not, *ceteris paribus*, lead to other assets increasing in price. Assuming that the economy is closed this assumption may be reasonable, as investors must make a decision as to how to partition their portfolios, and an increase in one partition should not increase other partitions.

## 6. *F*-expectations.

One interpretation of the solution to a BSDE is as a type of generalized expectation. In particular, for a fixed standard driver, $F$, and $Q \in \mathbb{R}^K$, an $\mathcal{F}_t$ measurable, square integrable random variable, we can define the conditional $F$-evaluation to be

$$\mathcal{E}_{s,t}^F(Q) = Y_s \tag{6.1}$$

for $s \leq t$ where $Y_s$ is the solution to

$$Y_s - \int_{]s,t]} F(\omega, u, Y_{u-}, Z_u)\, du + \int_{]s,t]} Z_u\, dM_u = Q.$$

DEFINITION 6.1. Following [15], we shall call a system of operators

$$\mathcal{E}_{s,t} \colon L^2(\mathcal{F}_t) \to L^2(\mathcal{F}_s), \qquad 0 \leq s \leq t \leq T,$$

an $\mathcal{F}_t$-consistent *nonlinear evaluation* for $\{\mathcal{Q}_{s,t} \subset L^2(\mathcal{F}_t) | 0 \leq s \leq t \leq T\}$ defined on $[0, T]$ if it satisfies the following properties:

1. For $Q, Q' \in \mathcal{Q}_{s,t}$,

$$\mathcal{E}_{s,t}(Q) \geq \mathcal{E}_{s,t}(Q')$$

   $\mathbb{P}$-a.s. componentwise whenever $Q \geq Q'$ $\mathbb{P}$-a.s. componentwise, with equality iff $Q = Q'$ $\mathbb{P}$-a.s.;
2. $\mathcal{E}_{t,t}(Q) = Q$ $\mathbb{P}$-a.s.;
3. $\mathcal{E}_{r,s}(\mathcal{E}_{s,t}(Q)) = \mathcal{E}_{r,t}(Q)$ $\mathbb{P}$-a.s. for any $r \leq s \leq t$;
4. For any $A \in \mathcal{F}_s$, $I_A \mathcal{E}_{s,t}(Q) = I_A \mathcal{E}_{s,t}(I_A Q)$ $\mathbb{P}$-a.s.

We are here allowing a significant generalization of [15], as our evaluations can all be vector valued, provided they are square integrable. This generalization allows the use of these evaluations in multi-objective problems, where a scalar evaluation process may be insufficient.

We also only require these properties to hold on some subset $\mathcal{Q}_{s,t}$ of the set of square integrable terminal conditions, a distinction which shall become important when we consider arbitrage and finite market modelling. Note, however, that $\mathcal{E}_{s,t}$ is defined over the whole of $L^2(\mathcal{F}_t)$, but it is only on $\mathcal{Q}_s^t$ that property 1 holds. Furthermore, we add the restriction that if $Q \geq Q'$ $\mathbb{P}$-a.s., $Q, Q' \in \mathcal{Q}_t$, then $\mathcal{E}_{s,t}(Q) = \mathcal{E}_{s,t}(Q')$ $\mathbb{P}$-a.s. iff $Q = Q'$ $\mathbb{P}$-a.s.

REMARK 6.1.   If $\mathcal{E}_{s,t}$ is an $\mathcal{F}_t$-consistent nonlinear evaluation for $\mathcal{Q}_{s,t}$, it is also an $\mathcal{F}_t$-consistent nonlinear evaluation for any subset of $\mathcal{Q}_{s,t}$.

DEFINITION 6.2.   We often wish for the sets $\{\mathcal{Q}_{s,t}\}$ to be stable through time. That is, $\mathcal{E}$ will be called *dynamically monotone* for $\{\mathcal{Q}_{s,t}\}$ if and only if, for all $r \le s \le t$:

   (i) $\mathcal{E}$ is an $\mathcal{F}_t$ consistent nonlinear evaluation for $\{\mathcal{Q}_{s,t}\}$;
   (ii) $\mathcal{Q}_{r,t} \subseteq \mathcal{Q}_{s,t}$ ($\mathcal{Q}_{s,t}$ is *nondecreasing* in $s$);
   (iii) $\mathcal{E}_{s,t}(Q) \in \mathcal{Q}_{r,s}$ for all $Q \in \mathcal{Q}_{r,t}$.

REMARK 6.2.   In an economic context, where $\mathcal{E}_{s,t}$ represents the price at $s$ of an asset to be sold at time $t$, dynamic monotonicity corresponds, in some sense, to the statements:

   (ii) "An asset with an arbitrage free price, when bought at a time $r$ and sold at time $t$, should also have an arbitrage free price when bought at any time $s$ following $r$ (with $s$ prior to $t$)" and

   (iii) "An asset with an arbitrage free price, when the asset is bought at a time $r$ and sold at time $t$, should be arbitrage free when sold at any time $s$ prior to $t$ (with $s$ following $r$)."

THEOREM 6.3.   *Fix a driver $F$. Consider a collection of sets $\{\mathcal{Q}_{s,t} \subset L^2(\mathcal{F}_t)\}$ with $\mathcal{Q}_{r,t} \subseteq \mathcal{Q}_{s,t}$ for all $r \le s \le t$. Suppose that, for any $Q^1, Q^2 \in \mathcal{Q}_{s,t}$, at least one of Theorems 4.2, 5.1, 5.3 and 5.7 holds on $[s,t]$, with $F^1 = F^2 = F$ whenever $Q^1 \ge Q^2$ $\mathbb{P}$-a.s. Then $\mathcal{E}_{s,t}^F$ defined in (6.1) is an $\mathcal{F}_t$-consistent nonlinear evaluation for $\{\mathcal{Q}_{s,t}\}$.*

PROOF.   1. The statement $\mathcal{E}_{s,t}^F(Q_1) \ge \mathcal{E}_{s,t}^F(Q_2)$ $\mathbb{P}$-a.s. whenever $Q_1 \ge Q_2$ $\mathbb{P}$-a.s. is simply the main result of each comparison theorem, one of which holds by assumption. The strict comparison then establishes the second statement.

2. The fact $\mathcal{E}_{t,t}^F(Q) = Q$ $\mathbb{P}$-a.s. for any $\mathcal{F}_t$ measurable $Q$ is trivial, as we have defined $\mathcal{E}_{t,t}^F(Q)$ by the solution to a BSDE, which reaches its terminal value $Q$ at time $t$ by construction.

3. For any $r \le s \le t$, let $Y$ denote the solution to the relevant BSDE. Then we have

$$Q = Y_r - \int_{]r,t]} F(\omega, u, Y_{u-}, Z_u)\, du + \int_{]r,t]} Z_u\, dM_u,$$

which implies

$$Y_s = Y_r - \int_{]r,s]} F(\omega, u, Y_{u-}, Z_u)\, du + \int_{]r,s]} Z_u\, dM_u.$$

Hence $Y_r$ is also the time $r$ value of a solution to the BSDE with terminal time $s$ and value $Y_s$. Hence

$$\mathcal{E}_{r,s}^F(\mathcal{E}_{s,t}^F(Q)) = \mathcal{E}_{r,t}^F(Q)$$

$\mathbb{P}$-a.s., as desired.

4. We wish to show that for $A \in \mathcal{F}_s$, $I_A \mathcal{E}_{s,t}(Q) = I_A \mathcal{E}_{s,t}(I_A Q)$ $\mathbb{P}$-a.s. Write $\mathcal{E}_{s,t}(Q)$ as the solution to the BSDE with terminal value $Q$, and $\mathcal{E}_{s,t}(I_A Q)$ as the solution to the BSDE with terminal value $I_A Q$. Premultiplying these BSDEs by $I_A$ gives two BSDEs, both with terminal value $I_A Q$, driver $\tilde{F}(\omega, t, Y_t, Z_t) = I_A F(\omega, t, Y_t, Z_t)$ and solutions $I_A \mathcal{E}_{s,t}(Q)$ and $I_A \mathcal{E}_{s,t}(I_A Q)$, respectively. From Theorem 1.1, the solution to this BSDE is unique, hence $I_A \mathcal{E}_{s,t}(Q) = I_A \mathcal{E}_{s,t}(I_A Q)$ up to indistinguishability. □

REMARK 6.4. As we have seen in Section 4.1, even in the scalar case, the assumption that a comparison theorem holds, and hence $\mathcal{E}_{s,t}^F$ is a nonlinear evaluation on $\mathcal{Q}_{s,t}$, is nontrivial.

DEFINITION 6.3. Let $\{\mathcal{Q}_{s,t} \subset L^2(\mathcal{F}_t)\}$ be a family of sets which are nondecreasing in $s$. For a BSDE driver $F^1$, suppose that, for all $s \leq t \leq T$, assumption (iii) [and assumption (iv), when applicable], of at least one of Theorems 4.2, 5.1, 5.3 and 5.7 hold on $]s, t]$ for all $Q^1, Q^2 \in \mathcal{Q}_{s,t}$ [whether or not assumption (i) holds]. Then $F^1$ is said to be a *balanced* driver on $\{\mathcal{Q}_{s,t}\}$.

REMARK 6.5. The logic of this name is due to the geometry of the problem, as, in some sense, $F^1$ here balances the outcomes with zero hedging within the range of outcomes with hedging.

LEMMA 6.6. *Let $F$ be a balanced driver on $\{\mathcal{Q}_{s,t}\}$ where $\mathcal{Q}_{s,t}$ is nondecreasing in $s$. Then $\mathcal{E}^F$ is an $\mathcal{F}_t$ consistent nonlinear evaluation on $\{\mathcal{Q}_{s,t}\}$.*

PROOF. The requirements for $F$ to be balanced are stronger than those needed for Theorem 6.3, and so the result follows. □

LEMMA 6.7. *Fix a driver $F$ balanced on $\{\mathcal{Q}_{s,t}\}$ where $\mathcal{Q}_{s,t}$ is nondecreasing in $s$. Then there exists a family of sets, $\{\tilde{\mathcal{Q}}_{s,t}\}$, such that $\mathcal{E}^F$ is dynamically monotone for $\{\tilde{\mathcal{Q}}_{s,t}\}$ and $\mathcal{Q}_{s,t} \subseteq \tilde{\mathcal{Q}}_{s,t}$ for all $s \leq t \leq T$.*

PROOF. Just as the proof of Theorem 6.3 (point 3), it is easy to see that $\mathcal{E}^F$ is recursive, that is,

$$\mathcal{E}_{r,s}^F(\mathcal{E}_{s,t}^F(Q)) = \mathcal{E}_{r,t}^F(Q).$$

Define

$$\tilde{\mathcal{Q}}_{s,t} = \{\mathcal{E}_{t,u}^F(Q) | Q \in \mathcal{Q}_{s,u} \text{ for some } u \geq t\}.$$

As $F$ is balanced on $\{\mathcal{Q}_{s,t}\}$, assumptions (iii) and (iv) of the relevant comparison theorem must hold on $]s, u]$ for all $Q \in \mathcal{Q}_{s,u}$. As $\mathcal{E}^F$ is recursive, it follows that assumptions (iii) and (iv) of the comparison theorem must hold on $]s, t]$ for all $Q \in \tilde{\mathcal{Q}}_{s,t}$, that is, $F$ is balanced on $\{\tilde{\mathcal{Q}}_{s,t}\}$. It follows from Lemma 6.3 that $\mathcal{E}^F$ is an $\mathcal{F}_t$ consistent nonlinear evaluation on $\{\tilde{\mathcal{Q}}_{s,t}\}$.

As $\mathcal{Q}_{s,t}$ is nondecreasing in $s$, it is clear that $\tilde{\mathcal{Q}}_{s,t}$ is nondecreasing in $s$. Also, as $\mathcal{E}_{t,t}^F(Q) = Q$, it is clear that $\mathcal{Q}_{s,t} \subseteq \tilde{\mathcal{Q}}_{s,t}$ for all $s \leq t$.

Finally, for any $Q \in \tilde{\mathcal{Q}}_{r,t}$, there exists a $u \geq t$ with $Q = \mathcal{E}_{t,u}^F(Q')$ for some $Q' \in \mathcal{Q}_{r,u}$. Therefore, by recursivity,

$$\mathcal{E}_{s,t}^F(Q) = \mathcal{E}_{s,u}^F(Q') \in \tilde{\mathcal{Q}}_{r,s}. \qquad \square$$

DEFINITION 6.4. Again following [15], we shall call a system of operators,

$$\mathcal{E}(\cdot|\mathcal{F}_t) : L^2(\mathcal{F}_T) \to L^2(\mathcal{F}_t), \qquad 0 \leq t \leq T,$$

an $\mathcal{F}_t$-consistent *nonlinear expectation* for $\{\mathcal{Q}_t \subset L^2(\mathcal{F}_T)\}$ defined on $[0, T]$ if it satisfies the following properties.

1. For $Q, Q' \in \mathcal{Q}_t$,

$$\mathcal{E}(Q|\mathcal{F}_t) \geq \mathcal{E}(Q'|\mathcal{F}_t)$$

   $\mathbb{P}$-a.s. componentwise whenever $Q \geq Q'$ $\mathbb{P}$-a.s. componentwise, with equality iff $Q = Q'$ $\mathbb{P}$-a.s.
2. $\mathcal{E}(Q|\mathcal{F}_t) = Q$ $\mathbb{P}$-a.s. for any $\mathcal{F}_t$ measurable $Q$.
3. $\mathcal{E}(\mathcal{E}(Q|\mathcal{F}_t)|\mathcal{F}_s) = \mathcal{E}(Q|\mathcal{F}_s)$ $\mathbb{P}$-a.s. for any $s \leq t$.
4. For any $A \in \mathcal{F}_t$, $I_A \mathcal{E}(Q|\mathcal{F}_t) = \mathcal{E}(I_A Q|\mathcal{F}_t)$ $\mathbb{P}$-a.s.

REMARK 6.8. It is clear that any nonlinear expectation is also a nonlinear evaluation, with $\mathcal{E}_{s,t}(\cdot) = \mathcal{E}(\cdot|\mathcal{F}_s)$ for all $s \leq t$, and hence the concept of dynamic monotonicity extends to this new setting. On the other hand, as the terminal time is irrelevant in this context, one can simply specify the sets $\mathcal{Q}_t = \mathcal{Q}_t^T$, as the expectation refers without loss of generality to the terminal values at time $T$. Therefore, we say $\mathcal{E}(\cdot|\mathcal{F}_t)$ is *dynamically monotone* for $\{\mathcal{Q}_t\}$ if, for all $s \leq t$:

  (i) $\mathcal{E}$ is an $\mathcal{F}_t$ consistent nonlinear expectation for $\{\mathcal{Q}_t\}$;
  (ii) $\mathcal{Q}_s \subseteq \mathcal{Q}_t$ ($\mathcal{Q}_t$ is *nondecreasing* in $t$);
  (iii) $\mathcal{E}(Q|\mathcal{F}_t) \in \mathcal{Q}_s$ for all $Q \in \mathcal{Q}_s$.

THEOREM 6.9. *Fix a driver $F$ such that $F(\omega, t, Y_{t-}, 0) = 0$ $dt \times \mathbb{P}$-a.s. Consider a family of sets, $\{\mathcal{Q}_t \subset L^2(\mathcal{F}_T)\}$, such that for any $Q, Q' \in \mathcal{Q}_t$ with* $Q \geq Q'$, at least one of Theorems *4.2*, *5.1*, *5.3* and *5.7* holds on $]t, T]$ with $F^1 = F^2 = F$. The functional $\mathcal{E}^F(\cdot | \mathcal{F}_t)$ defined for each $t$ by

(6.2) $$\mathcal{E}^F(Q | \mathcal{F}_t) = Y_t,$$

where $Y_t$ is the solution to

$$Y_t - \int_{]t, T]} F(\omega, u, Y_{u-}, Z_u) \, du + \int_{]t, T]} Z_u \, dM_u = Q$$

is an $\mathcal{F}_t$-consistent nonlinear expectation for $\{\mathcal{Q}_t\}$.

PROOF. Properties 1 and 3 follow exactly as in the proof of Theorem *6.3*.

Property 2 follows because we know that for $t < T$, if $Q$ is $\mathcal{F}_t$ measurable then the solution of

$$Y_t - \int_{]t, T]} F(\omega, u, Y_{u-}, Z_u) \, du + \int_{]t, T]} Z_u \, dM_u = Q$$

has $Z_u = 0 \ d\langle M, M \rangle_u \times \mathbb{P}$-a.s. (Simply take an $\mathcal{F}_t$ conditional expectation as done earlier.) Hence for $t \leq u \leq T$, $F(\omega, u, Y_{u-}, Z_u) = 0$, and therefore $Y_t = Q$ $\mathbb{P}$-a.s., as desired.

For property 3, we know that $I_A \mathcal{E}(Q | \mathcal{F}_t)$ is the solution to a BSDE with terminal value $I_A Q$ and driver $I_A F(\omega, t, Y, Z)$. As $F(\omega, t, Y_{t-}, 0) = 0$, $I_A F(\omega, t, Y, Z) = F(\omega, t, Y, I_A Z)$ $\mathbb{P}$-a.s. We also know $\mathcal{E}(I_A Q | \mathcal{F}_t)$ is the solution to a BSDE with driver $F$, and taking an $\mathcal{F}_t$ conditional expectation shows that the solution $Z$ process for $\mathcal{E}(I_A Q | \mathcal{F}_t)$ satisfies $Z = I_A Z$. Hence the two quantities solve the same BSDE, and so by the uniqueness of Theorem *1.1* must be equal. □

COROLLARY 6.10. *Fix a driver $F$ balanced on $\{\mathcal{Q}_t\}$ where $\mathcal{Q}_t$ is nondecreasing in $t$ and $F(\omega, t, Y_{t-}, 0) = 0 \ dt \times \mathbb{P}$-a.s. Then there exists a family of sets $\{\tilde{\mathcal{Q}}_t\}$ such that $\mathcal{E}^F$ is dynamically monotone for $\{\tilde{\mathcal{Q}}_t\}$ and $\mathcal{Q}_t \subseteq \tilde{\mathcal{Q}}_t$ for all $s \leq t \leq T$.*

PROOF. This follows exactly as in Lemmas *6.6* and *6.7* and Theorem *6.9*. □

THEOREM 6.11. *For any $\{\mathcal{Q}_t\}$, there exists a nonlinear $F$-expectation on $\{\mathcal{Q}_t\}$.*

PROOF. Simply take $F \equiv 0$. Then the nonlinear expectation corresponds to the classical expectation (and is balanced on $\{\mathcal{Q}_t\}$), and immediately satisfies all the desired properties. □

**7. Dominance, arbitrage and $F$-expectations.** In this section we shall explore in detail the behavior of $\mathcal{E}^F$ given different sets $\{\mathcal{Q}_{s,t}\}$. We wish to discuss the relationship between $F$-expectations, arbitrage and assumption (iii) of Theorem 4.2. This will highlight how this model can be used to understand phenomena which may be lost under a model driven by Brownian motion and the importance of considering nonlinear expectations only for subsets $\{\mathcal{Q}_{s,t} \subset L^2(\mathcal{F}_t)\}$. Throughout this section, we will consider only a single, fixed driver $F$, and all vector inequalities are to be read componentwise.

In this context we need to distinguish between the closely related concepts of arbitrage and dominance. This is essentially because our "pricing rule," as defined by the solutions to a BSDE, is nonlinear.

We begin with definitions of dominance and arbitrage in this context.

DEFINITION 7.1. Fix an initial time $s$. Consider a terminal time $t \geq s$ and a pair of terminal values $Q^1, Q^2 \in L^2(\mathcal{F}_t)$. Let $Y^1$ and $Y^2$ a pair of corresponding evaluation processes. If, for some $A \in \mathcal{F}_s$, we have $Q^1 \geq Q^2$ and $Y_s^1 \leq Y_s^2$ both $\mathbb{P}$-a.s. on $A$, and at least one of these inequalities is strict with positive probability on $A$, then we shall say that $Q^1$ *dominates* $Q^2$ at $s$, under this evaluation, given $A$.

DEFINITION 7.2. Fix an initial time $s$. Consider a terminal time $t \geq s$ and terminal value $Q \in L^2(\mathcal{F}_t)$. Let $Y$ be the the corresponding evaluation process. If, for some $A \in \mathcal{F}_s$, we have $Q \geq 0$ and $Y_s \leq 0$, both $\mathbb{P}$-a.s. on $A$, and at least one of these inequalities is strict with positive probability on $A$, then we shall say that $Q$ is an *arbitrage opportunity* at $s$, under this evaluation, given $A$.

DEFINITION 7.3. We shall say that a set $\mathcal{Q}_{s,t} \subset L^2(\mathcal{F}_t)$ allows dominance if there exists $Q^1, Q^2 \in \mathcal{Q}_{s,t}$ with $Q^1$ dominating $Q^2$ at $s$. Similarly for arbitrage.

REMARK 7.1. If the evaluation of the terminal value $Q = 0$ is $Y_s = 0$ $\mathbb{P}$-a.s. on $A$, then an arbitrage opportunity is simply a strategy which dominates the zero strategy.

This distinction and terminology has a simple interpretation when we think of our evaluations as prices. An "arbitrage" opportunity is any situation where, for no positive outlay today, a positive outcome can be ensured in the future. On the other hand, for an asset to dominate another, is to state a strict preference relation between the two, as the better option in the future is at least as cheap today. (Examples of this price interpretation of BSDE solutions, for example, to option pricing under constraints, can be found in [7].)

It is significant in this context that these situations are not scale-invariant. This means that there may be a situation where there is a small arbitrage opportunity, but the nonlinearity of the pricing rule means that this cannot be replicated significantly.

When $F$ is linear and $\mathcal{Q}_{s,t}$ is a vector space, as is implicitly assumed in a Black–Scholes market model, the existence of dominance and the existence of arbitrage are equivalent. In our context, we need to maintain the distinction between them, particularly due to the nonlinearity of $F$.

THEOREM 7.2. *For any $\mathcal{F}_t$ consistent, nonlinear evaluation for $\{\mathcal{Q}_{s,t}\}$ (in the sense of Definition 6.1), the sets $\mathcal{Q}_{s,t}$ do not allow dominance under this evaluation.*

PROOF. This is simply the first property of nonlinear evaluations. □

COROLLARY 7.3. *Let $F$ be a balanced driver on $\{\mathcal{Q}_{s,t}\}$. Then the sets $\mathcal{Q}_{s,t}$ do not allow dominance, under the pricing rule given by the BSDE solutions with driver $F$.*

PROOF. As $F$ is balanced, by Lemma 6.6 we know that such an $F$ defines an $\mathcal{F}_t$ consistent nonlinear evaluation for $\{\mathcal{Q}_{s,t}\}$. □

THEOREM 7.4. *For any $\mathcal{F}_t$ consistent nonlinear expectation for $\{\mathcal{Q}_t\}$ with $0 \in \mathcal{Q}_t$ for all $t$, the sets $\mathcal{Q}_t$ do not allow either arbitrage or dominance under this nonlinear expectation.*

PROOF. These are simply the first and second properties of nonlinear expectations, and an application of Remark 7.1. □

COROLLARY 7.5. *Consider a driver $F$ satisfying $F(\omega, t, 0, 0) = 0$, $dt \times \mathbb{P}$-a.s. on $[0, T]$. Then if $F$ is a balanced driver on some sets $\{\mathcal{Q}_t\}$ with $0 \in \mathcal{Q}_t$ for all $t$, the sets $\mathcal{Q}_t$ do not allow arbitrage under the evaluation given by $\mathcal{E}^F$.*

PROOF. We know that such an $F$ defines a nonlinear expectation on $\mathcal{Q}$, by Corollary 6.10. □

As we have seen in Section 4.1, there exist examples of BSDE solutions in this context where arbitrage opportunities do exist. This is a significant distinction between these models based on Markov chains and models based on Brownian motions.

REMARK 7.6. For given market data, we may ask when the data are consistent with arbitrage free BSDE pricing. Let $\mathcal{Q}_{0,t}$ denote the time $t$ values of all possible tradable assets in the market, including combinations of assets. Our question is then equivalent to asking if there exists a driver $F$ such that there is a nonlinear evaluation $\mathcal{E}_{0,t}^F(\cdot)$ for $\{\mathcal{Q}_{0,t}\}$ which maps each terminal condition to its current price.

If there is a unique $F$ that will do this, we have a complete, arbitrage free, BSDE pricing scheme. This does not guarantee that the BSDE solutions for terminal conditions not in $\{\mathcal{Q}_{0,t}\}$ will be valid, arbitrage free prices.

REMARK 7.7. Under BSDE pricing with Markov chains, we may have arbitrage free pricing for those assets in the market, but not for other assets we wish to consider. If we were to hypothetically introduce other assets into the market, while retaining no arbitrage, this could result in changing the driver $F$, and hence the dynamics of all stocks in the market.

**8. Geometry.** Different assumptions on $\{\mathcal{Q}_{s,t}\}$ and $F$ imply different geometric results on the values of $Y_t$. For simplicity, we here restrict our attention to the scalar $(K = 1)$ case.

DEFINITION 8.1. For any $s$, we define $H_s(Q)$ to be the *essential convex hull* of $Q$ at time $s$ to be the smallest $\mathcal{F}_s$ measurable convex set such that $P(Q \in H_s(Q)|\mathcal{F}_s) = 1$.

DEFINITION 8.2. We define r.i. $H_s(Q)$ to be the relative interior of $H_s(Q)$, that is, the interior of $H_s(Q)$ viewed as a subset of the affine hull it generates.

REMARK 8.1. The interested reader is referred to any good book on elementary stochastic finance for a more detailed definition (e.g., [11], page 27 or [10], page 65).

THEOREM 8.2. *Consider a scalar $(K = 1)$, nonlinear $F$-expectation $\mathcal{E}^F(\cdot|\mathcal{F}_t)$ for $\{\mathcal{Q}_t\}$. Suppose, for all $q \in \mathbb{R}$, $q \in \mathcal{Q}_0$. Then for all $Q \in \mathcal{Q}_0$, $\mathbb{P}$-a.s.,*

$$\mathcal{E}^F(Q|\mathcal{F}_0) \in \text{r.i. } H_0(Q).$$

PROOF. If $H_0(Q)$ is singular, that is, it contains only a single point, then $Q$ is $\mathcal{F}_0$ measurable (up to equality $\mathbb{P}$-a.s.), and so $Y_0 = Q$ $\mathbb{P}$-a.s., and the condition is trivial.

Otherwise, we first wish to show that $\mathcal{E}^F(Q|\mathcal{F}_0) > \inf H_0(Q)$. Let

$$Q^{\min} = \inf H_0(Q).$$

As this is $\mathcal{F}_0$ measurable, the solution to the BSDE with driver $F$ and terminal condition $Q^{\min}$ is simply $Y_0 = Q^{\min}$. If $H_0(Q)$ is unbounded below, $Q^{\min} = -\infty$ and the statement is trivial. Otherwise, $Q^{\min} \in \mathbb{R}$, and so $Q^{\min} \in \mathcal{Q}$. No dominance in $\mathcal{Q}$ then implies that

$$\mathcal{E}^F(Q|\mathcal{F}_0) > Q^{\min}.$$

We can then repeat this argument with $Q^{\max} = \sup H_0(Q)$, to show that $\mathcal{E}^F(Q|\mathcal{F}_0) < \sup H_0(Q)$. Hence $\mathcal{E}^F(Q|\mathcal{F}_0)$ lies strictly within the interior of $H_0(Q)$, which, as we are dealing with nonsingular, scalar valued $Q$, corresponds precisely with the set r.i. $H_0(Q)$ as desired. $\square$

REMARK 8.3. Clearly, this result can be extended to show $\mathcal{E}^F(Q|\mathcal{F}_t) \in$ r.i. $H_t(Q)$. However, care must be taken as there is no guarantee that $Q^{\min} := \inf H_t(Q) \in L^2(\mathcal{F}_T)$, even though $Q^{\min}$ is $\mathcal{F}_t$ measurable.

To show this general result, the existence theorem must be strengthened to show that if $E[Q^2|\mathcal{F}_t] < +\infty$ $\mathbb{P}$-a.s., then there exists a solution $(Y, Z)$ to the BSDE (1.3) on $[t, T]$. With this extension, we can weaken our definition of $\mathcal{Q}_t$ to be

$$\mathcal{Q}_t \subseteq \{Q : Q \in L^0(\mathcal{F}_T), E[Q^2|\mathcal{F}_t] < +\infty, \mathbb{P}\text{-a.s.}\}.$$

We then assume that $L^0(\mathcal{F}_t) \subseteq \mathcal{Q}_t$, and the result follows.

## 9. Other properties of $\mathcal{E}^F$ and applications to risk measures.
One use of BSDEs which has developed recently is as a framework for developing dynamically consistent, convex or coherent risk measures. This can be seen in [2], and more generally in [16]. We here present the key results in the context of BSDEs driven by Markov chains.

As noted in [15], the theory of nonlinear expectations provides an ideal setting for the discussion of risk measures in the sense of [1] and others. We shall consider dynamic risk measures $\rho_t^F$ of the form

$$\rho_t^F(Q) := -\mathcal{E}_{t,T}^F(Q),$$

where $\mathcal{E}^F$ is a nonlinear $F$-evaluation for some $\{\mathcal{Q}_{s,T} \subset L^2(\mathcal{F}_T)\}$. As before, our approach allows a considerable generalization of earlier results, as the quantities considered can be vector valued.

An alternative specification, used in [2] and [16], is to let $\rho_t^F(Q) = \mathcal{E}_{t,T}^F(-Q)$, the solution to a BSDE with terminal condition $-Q$. Under this alternative specification, the following results remain valid, however, the result of Theorems 9.7 holds for $\rho$ with $F$ convex rather than concave.

We seek to determine properties of $\rho_t^F$, or, equivalently, of $\mathcal{E}_{t,T}^F$ which are particularly relevant in the context of risk measures.

The first four properties below follow directly from the properties of nonlinear evaluations, recalling the proofs of Theorems 6.3 and 6.9, with $\mathcal{Q}_t = \varnothing$ where appropriate.

LEMMA 9.1 (Terminal equality). *For any $F$, $\rho_t^F$ satisfies the terminal condition $\rho_T^F(Q) = -Q$.*

PROOF. Equivalent to property 2 of nonlinear evaluations.  □

THEOREM 9.2 (Dynamic consistency). *For any $F$, $\rho_t^F$ satisfies the recursivity and dynamic consistency requirements,*

$$\rho_s^F(Q) = \rho_s^F(-\rho_t^F(Q))$$

*and*

$$\rho_t^F(Q) = \rho_t^F(Q') \quad \Rightarrow \quad \rho_s^F(Q) = \rho_s^F(Q'),$$

*for times $s \le t \le T$, all equalities being $\mathbb{P}$-a.s.*

PROOF. The first statement is equivalent to property 3 of nonlinear evaluations. The second follows by uniqueness of solutions to BSDEs with given terminal conditions, as in Theorem 1.1.  □

LEMMA 9.3 (Constants). *Let $Q$ be $\mathbb{P}$-a.s. equal to a $\mathcal{F}_t$ measurable terminal condition. Then if $F$ is a driver with normalization $F(\omega, t, Q, 0) = 0$ $dt \times \mathbb{P}$-a.s., $\rho_t^F(Q) = -Q$ for all $t$.*

PROOF. Equivalent to property 4 of nonlinear expectations.  □

THEOREM 9.4 (Monotonicity). *Let $F$ be a balanced driver for $\{\mathcal{Q}_{t,T}\}$. Then for any $Q^1, Q^2 \in \mathcal{Q}_{t,T}$ with $Q^1 \ge Q^2$ $\mathbb{P}$-a.s., we have $\rho_t^F(Q^1) \le \rho_t^F(Q^2)$, with equality if and only if $Q^1 = Q^2$, $\mathbb{P}$-a.s.*

PROOF. Equivalent to property 1 of nonlinear evaluations.  □

LEMMA 9.5 (Translation invariance). *Let $F$ be a driver for a BSDE with normalization $F(\omega, u, Y_{u-}, 0) = 0$ $\mathbb{P} \times du$-a.s., on $]t, T]$. Suppose $F$ does not depend on $Y_{u-}$. Then for any $q \in L^2(\mathcal{F}_t)$,*

$$\rho_t^F(Q + q) = \rho_t^F(Q) - q,$$
$$\mathcal{E}_{t,T}^F(Q + q) = \mathcal{E}_{t,T}^F(Q) + q.$$

PROOF. As $F$ does not depend on $Y_{u-}$, we know that

$$Y_t - \int_{]t,T]} F(\omega, u, Y_{u-}, Z_u) \, du + \int_{]t,T]} Z_u \, dM_u = Q$$

implies

$$[Y_t + q] - \int_{]t,T]} F(\omega, u, [Y_{u-} + q], Z_u) \, du + \int_{]t,T]} Z_u \, dM_u = [Q + q].$$

The result follows.  □

THEOREM 9.6 (Positive homogeneity). *Let $F$ be a positively homogeneous driver, that is, for any $\lambda > 0$, if $(Y, Z)$ is the solution corresponding to some $Q$, $\mathbb{P} \times du$-a.s. on $]t, T]$,*

$$F(\omega, u, \lambda Y_{u-}, \lambda Z_u) = \lambda F(\omega, u, Y_{u-}, Z_u).$$

*Then for all such $Q$, $\rho_t^F(\lambda Q) = \lambda \rho_t^F(Q)$ and $\mathcal{E}_{t,T}^F(\lambda Q) = \lambda \mathcal{E}_{t,T}^F(Q)$.*

PROOF. Simply multiply the BSDE with terminal condition $Q$ through by $\lambda$. $\square$

THEOREM 9.7 (Convexity/concavity). *Suppose $\mathcal{Q}_{t,T}$ is a convex set. Let $F$ be a concave balanced driver for $\mathcal{Q}_{t,T}$, that is, for any $\lambda \in [0, 1]$, any $(Y^1, Z^1), (Y^2, Z^2)$ the solutions corresponding to $Q^1, Q^2 \in \mathcal{Q}_{t,T}$, $\mathbb{P} \times du$-a.s. on $]t, T]$,*

$$F(\omega, u, \lambda Y_{u-}^1 + (1 - \lambda) Y_{u-}^2, \lambda Z_u^1 + (1 - \lambda) Z_u^2)$$

$$\geq \lambda F(\omega, u, Y_{u-}^1, Z_u) + (1 - \lambda) F(\omega, u, Y_{u-}^2, Z_u^2).$$

*Then for any $\lambda \in [0, 1]$ and any $Q^1, Q^2 \in \mathcal{Q}_{t,T}$,*

$$\rho_t^F(\lambda Q^1 + (1 - \lambda) Q^2) \leq \lambda \rho_t^F(Q^1) + (1 - \lambda) \rho_t^F(Q^2),$$

$$\mathcal{E}_{t,T}^F(\lambda Q^1 + (1 - \lambda) Q^2) \geq \lambda \mathcal{E}_{t,T}^F(Q^1) + (1 - \lambda) \mathcal{E}_{t,T}^F(Q^2).$$

PROOF. Taking a convex combination of the BSDEs with terminal conditions $Q^1$ and $Q^2$ gives the equation

$$\lambda Y_t^1 + (1 - \lambda) Y_t^2 - \int_{]t,T]} [\lambda F(\omega, u, Y_{u-}^1, Z_u) + (1 - \lambda) F(\omega, u, Y_{u-}^2, Z_u^2)] \, du$$

$$+ \int_{]t,T]} [\lambda Z_u^1 + (1 - \lambda) Z_u^2] \, dM_u = \lambda Q^1 + (1 - \lambda) Q^2,$$

which is a BSDE with terminal condition $\lambda Q^1 - (1 - \lambda) Q^2$ and driver

$$\tilde{F} = \lambda F(\omega, u, Y_{u-}^1, Z_u) + (1 - \lambda) F(\omega, u, Y_{u-}^2, Z_u^2).$$

We next consider the BSDE with terminal condition $\lambda Q^1 + (1 - \lambda) Q^2$ and driver $F$. Denote the solution to this by $Y^\lambda$. We can compare these BSDEs using the relevant comparison theorem—the first assumption of the theorem is trivial, the second is satisfied by the convexity of $F$, and the remainder because $F$ is balanced on $\mathcal{Q}_{t,T}$. Hence, our solutions satisfy

$$Y^\lambda \geq \lambda Y^1 + (1 - \lambda) Y^2$$

with equality if and only if the terminal conditions are equal with conditional probability one. The inequality for $\mathcal{E}_{t,T}^F$ follows. The inequality for $\rho_t^F$ can then be established by noting that as $\mathcal{E}_{t,T}^F(Q)$ is concave in $Q$, it follows that $\rho_t^F(Q) = -\mathcal{E}_{t,T}^F(Q)$ is convex in $Q$. $\square$

**10. Conclusions.** BSDEs with an underlying finite state Markov chain as a stochastic process form a viable alternative to BSDEs based on Brownian motions and Brownian motions with Poisson jumps. We have given closed-form solutions to linear BSDEs on Markov chains, and then have used these to develop a comparison theorem for the solutions.

These BSDEs allow for very delicate modelling of no arbitrage conditions. In particular, they allow no arbitrage only to be required on a subset of all possible terminal conditions, but the pricing rule to allow arbitrage outside this set. This is in contrast to BSDEs based on Brownian motion, where the comparison theorem guarantees no arbitrage in all cases. We have drawn a connection between the exclusion of arbitrage and the presence of a balanced driver $F$, a condition which can be understood as an infinitesimal no-arbitrage condition. We have also discussed the geometry of these BSDEs and the relation to no arbitrage conditions.

Finally, we have shown key properties which can be used when developing nonlinear evaluations and risk measures from these BSDEs. The added requirements of the theorem help us to distinguish precisely what is needed for each of these properties to hold which may prove useful when considering other models.

## APPENDIX: EXAMPLE PREDICTABLE QUADRATIC COVARIATION MATRIX

For clarity, an example of $\psi_t$ is presented here for $N = 4$. Suppose $X_{t-} = [0, 0, 1, 0]^*$, $A_t X_{t-} = [a_1, a_2, -a_1 - a_2 - a_4, a_4]^*$. Then

$$\psi_t = \begin{bmatrix} a_1 & 0 & -a_1 & 0 \\ 0 & a_2 & -a_2 & 0 \\ -a_1 & -a_2 & a_1 + a_2 + a_4 & -a_4 \\ 0 & 0 & -a_4 & a_4 \end{bmatrix}.$$

**Acknowledgments.** The authors would like to thank Charles Pearce, who read a copy of the manuscript. They also wish to express gratitude to an anonymous referee who read a previous version of the paper with exceptional care and diligence. The referee's corrections and suggestions have resulted in a greatly improved contribution.

Robert Elliott wishes to thank the SSHRC and the Australian Research Council for support.

## REFERENCES

[1] ARTZNER, P., DELBAEN, F., EBER, J.-M. and HEATH, D. (1999). Coherent measures of risk. *Math. Finance* **9** 203–228. MR1850791

[2] Barrieu, P. and El Karoui, N. (2004). Optimal derivatives design under dynamic risk measures. In *Mathematics of Finance* (G. Yin and Q. Zhang, eds.). *Contemporary Mathematics* **351** 13–25. Amer. Math. Soc., Providence, RI. MR2076287

[3] Bellman, R. (1970). *Introduction to Matrix Analysis*, 2nd ed. McGraw-Hill, New York. MR0258847

[4] Cohen, S. N. and Elliott, R. J. (2008). Solutions of backward stochastic differential equations on Markov chains. *Commun. Stoch. Anal.* **2** 251–262. MR2446692

[5] Cohen, S. N. and Elliott, R. J. (2009). A general theory of finite state backward stochastic difference equations. Forthcoming. Available at http://arxiv.org/abs/0810.4957v1.

[6] Cohen, S. N., Elliott, R. J. and Pearce, C. E. M. (2009). A ring isomorphism and corresponding pseudoinverses. Forthcoming. Available at http://arxiv.org/abs/0810.0093v1.

[7] El Karoui, N., Peng, S. and Quenez, M. C. (1997). Backward stochastic differential equations in finance. *Math. Finance* **7** 1–71. MR1434407

[8] Elliott, R. J. (1982). *Stochastic Calculus and Applications. Applications of Mathematics* **18**. Springer, New York. MR678919

[9] Elliott, R. J., Aggoun, L. and Moore, J. B. (1995). *Hidden Markov Models: Estimation and Control. Applications of Mathematics* **29**. Springer, New York. MR1323178

[10] Elliott, R. J. and Kopp, P. E. (2005). *Mathematics of Financial Markets*, 2nd ed. Springer, New York. MR2098795

[11] Föllmer, H. and Schied, A. (2002). *Stochastic Finance: An Introduction in Discrete Time. De Gruyter Studies in Mathematics* **27**. de Gruyter, Berlin. MR1925197

[12] Gantmacher, F. (1960). *Matrix Theory* **2**. Chelsea, New York.

[13] Gill, R. D. and Johansen, S. (1990). A survey of product-integration with a view toward application in survival analysis. *Ann. Statist.* **18** 1501–1555. MR1074422

[14] Jacod, J. and Shiryaev, A. N. (2003). *Limit Theorems for Stochastic Processes*, 2nd ed. Springer, Berlin. MR1943877

[15] Peng, S. (2005). Dynamically consistent nonlinear evaluations and expectations. Preprint No. 2004-1. Institute of Mathematics, Shandong Univ. Available at http://arxiv.org/abs/math/0501415.

[16] Rosazza Gianin, E. (2006). Risk measures via *g*-expectations. *Insurance Math. Econom.* **39** 19–34. MR2241848

School of Mathematical Sciences
University of Adelaide
SA 5005
Australia
E-mail: samuel.cohen@adelaide.edu.au
URL: http://www.maths.adelaide.edu.au/samuel.cohen/

School of Mathematical Sciences
University of Adelaide
SA 5005
Australia
and
Haskayne School of Business
University of Calgary
T2N 1N4
Canada
E-mail: robert.elliott@adelaide.edu.au
URL: http://www.ucalgary.ca/˜relliott